\documentclass[useAMS,usenatbib,referee]{biom}

\usepackage[ruled,vlined]{algorithm2e}
\usepackage{amsmath,mathabx}
\usepackage{graphicx,psfrag,epsf}
\usepackage{enumerate}
\usepackage{natbib}
\usepackage{url} % not crucial - just used below for the URL 
\usepackage{booktabs,threeparttable}
\usepackage{bbm}
\usepackage{bm}
\usepackage{mathtools}

\usepackage{etoolbox}
\usepackage[table]{xcolor}

\newcommand\revision[1]{{\color{black}#1}}
\newcommand\revisiontwo[1]{{\color{black}#1}}
\newcommand\revisionthree[1]{{\color{red}#1}}

\usepackage{bm}

\def\bff{{\bm{ f}}}
\def\bfg{{\bm{ g}}}
\def\bfh{{\bm{ h}}}

\def\bfn{{\bm{ n}}}

\def\bfD{{\bm{ D}}}

\def\bfS{{\bm{ S}}}

\def\bfV{{\bm{V}}}

\def\bfX{{\bm{ X}}}
\def\bfY{{\bm{ Y}}}
\def\bfZ{{\bm{ Z}}}
\def\bfzero{{\ensuremath{\bf 0}}}
\def\bfone{{\ensuremath{\bf 1}}}

\def\bftheta{{\ensuremath\boldsymbol{\theta}}}

\def\bfbeta{{\ensuremath\boldsymbol{\beta}}}

\def\bfepsilon{{\ensuremath\boldsymbol{\epsilon}}}

\def\bfvarOmega{{\ensuremath\boldsymbol{\varOmega}}}
\def\bfvarGamma{{\ensuremath\boldsymbol{\varGamma}}}

\def\bfvarSigma{{\ensuremath\boldsymbol{\varSigma}}}
\def\bfvarLambda{{\ensuremath\boldsymbol{\varLambda}}}

\def\bflambda{{\ensuremath\boldsymbol{\lambda}}}

\usepackage[figuresright]{rotating}

\title[An Efficient Data Integration Scheme]{An Efficient Data Integration Scheme for Synthesizing Information from Multiple Secondary {Datasets} for the \revision{Parameter Inference of the Main Analysis}}

\author{Chixiang Chen,$^{1*}$ Ming Wang,$^{2}$ Shuo Chen$^{1}$\\
	$^{1}$ Division of Biostatistics and Bioinformatics,\\ University of Maryland School of Medicine, Baltimore, MD, USA\\ 
	$^{2}$ \revision{Department of Population and Quantitative Health Sciences},\\ \revision{School of Medicine, Case Western Reserve University, Cleveland, OH, USA}\\ 
	{$^{*}$Contact Email: \url{chixiang.chen@som.umaryland.edu}}}

\begin{document}

\date{}

\pagerange{\pageref{firstpage}--\pageref{lastpage}} \pubyear{2023}

\volume{}
\artmonth{}
\doi{}

%  This label and the label ``lastpage'' are used by the \pagerange
%  command above to give the page range for the article

\label{firstpage}

%  pub the summary here

\begin{abstract}
{Many observational studies and clinical trials collect various secondary outcomes that may be highly correlated with the primary endpoint. These secondary outcomes are often analyzed in secondary analyses separately from the main data analysis. However, these secondary outcomes can be used to improve the estimation precision in the main analysis. We propose a method called Multiple Information Borrowing (MinBo) that borrows information from secondary data (containing secondary outcomes and covariates) to improve the efficiency of the main analysis. The proposed method is robust against model misspecification of the secondary data. Both theoretical and case studies demonstrate that MinBo outperforms existing methods in terms of efficiency gain. We apply MinBo to data from the Atherosclerosis Risk in Communities study to assess risk factors for hypertension.}
\end{abstract}


\maketitle

\section{Introduction}
\label{Introduction}
{In recent decades, statisticians have studied (semi-parametric) efficient estimators that fully utilize the main dataset \citep{shao2003mathematical, tsiatis2006semiparametric}. With the increasing availability of data from contemporary studies, the precision of estimation can be further improved by leveraging and integrating multiple sources of data \citep{qin2022selective}. Applications of data integration have been widely studied in many fields, such as epidemiology, clinical trials, genetics, and genomics \citep{lee2020estimation,li2020target}. In this paper, we introduce a new perspective and propose a novel framework that effectively synthesizes information from multiple sources of secondary data 
to improve the precision of parameter estimation in the main analysis model. Here, we regard a dataset as secondary if it contains a secondary outcome that is highly associated with the primary endpoint and some covariates.}

Our work is highly motivated by the Atherosclerosis Risks in Communities (ARIC) study \citep{aric1989atherosclerosis, gonzalez2018midlife}. Our goal is to identify risk factors for the development of hypertension, which affects over one-third of U.S.\ adults and is one of the leading causes of cardiovascular disease worldwide. Essential hypertension is defined as systolic blood pressure (SBP) $\geq 140$~mm~Hg or diastolic blood pressure (DBP) $\geq 90$~mm~Hg, or the need to take anti-hypertensive medication. Logistic regression with hypertension status as the primary endpoint has traditionally been used to investigate risk factors such as smoking, drinking, and age \citep{al2007hypertension}. However, this well-established maximum likelihood estimation (MLE) approach may not be efficient, as it does not incorporate information about SBP, DBP, or medication use in the main analysis. {Since these secondary outcomes are highly correlated with the primary endpoint, they are believed to carry additional information that could help improve parameter estimation in the main model.} Secondary outcomes are also common in other epidemiology studies and clinical trials. For instance, in risk prediction for stroke as the primary interest, other events, such as myocardial infarction and peripheral vascular disease, may be secondary outcomes and can provide information related to stroke \citep{ wilson2011coffee}. \revision{These real scenarios motivate us to develop a new data integration tool that can effectively synthesize information from multiple secondary 
datasets to improve parameter estimation.}

\revision{\revisiontwo{In the literature, most existing data-integration techniques are designed to incorporate information from external or independent studies.} For instance, the generalized meta-analysis \citep{kundu2019generalized} and constrained maximum likelihood approaches \citep{chatterjee2016constrained,han2017empirical} utilize summary information from external studies. Empirical likelihood estimation \citep{qin1994empirical,qin2000miscellanea,qin2015using} and calibration estimation \citep{lumley2011connections,yang2019combining} consider some type of shared or known effects across different studies. Bayesian integration \citep{xie2013incorporating,cheng2018improving} utilizes prior knowledge from external sources. \revisiontwo{However, these methods cannot be directly applied to our data framework as the primary endpoint is highly correlated with the secondary outcomes because they are from the same study.}} \revisiontwo{One potentially applicable approach is to specify a joint likelihood for all the outcomes by jointly modeling their means and correlations via likelihood. Although, the joint modeling of multiple outcomes is conceptually achievable, it is computationally complicated due to complexities in outcome features and real-world-data collection schemes. Moreover, the violation of any of the distribution assumptions among the outcomes will lead to inconsistent parameter estimators in the main model.} To circumvent these issues, \cite{chen2021improving,chen2022synthesizing} proposed a re-weighting estimation scheme that efficiently and robustly delivered information from a secondary dataset without the need for joint modeling. \revisiontwo{However, this approach is applicable only if there is only one secondary dataset. In practice, multiple secondary datasets are often available, and efficiently integrating the information may further improve parameter estimation in the main model.} 

We introduce here an efficient and robust scheme called \textbf{M}ultiple \textbf{in}formation \textbf{Bo}rrowing (named MinBo) that leverages various secondary outcomes. \revision{Using the proposed scheme, information can be effectively borrowed from multiple sources of information, resulting in improved estimation efficiency in the main analysis and good computational efficiency. The estimation consistency is guaranteed under a mild condition and is not sensitive to a misspecification of the ``working" model for any secondary data.} The rest of the manuscript is organized as follows. Section~\ref{Proposed Framework} describes the general estimation framework in MinBo. Section~\ref{Simulation Studies} presents extensive numerical evaluations. Section~\ref{An real data application} illustrates the application of MinBo to the ARIC study. The aim is to identify risk factors in the development of hypertension. Section~\ref{conclusion} concludes. Detailed proofs, additional numerical results, and a further discussion are included in the Supporting Information.

\section{The proposed framework: MinBo}
\label{Proposed Framework}
\subsection{Basic setup and existing methods}\label{Main and multiple auxiliary data}

\begin{figure}
    \centering
    \includegraphics[width=6.5in]{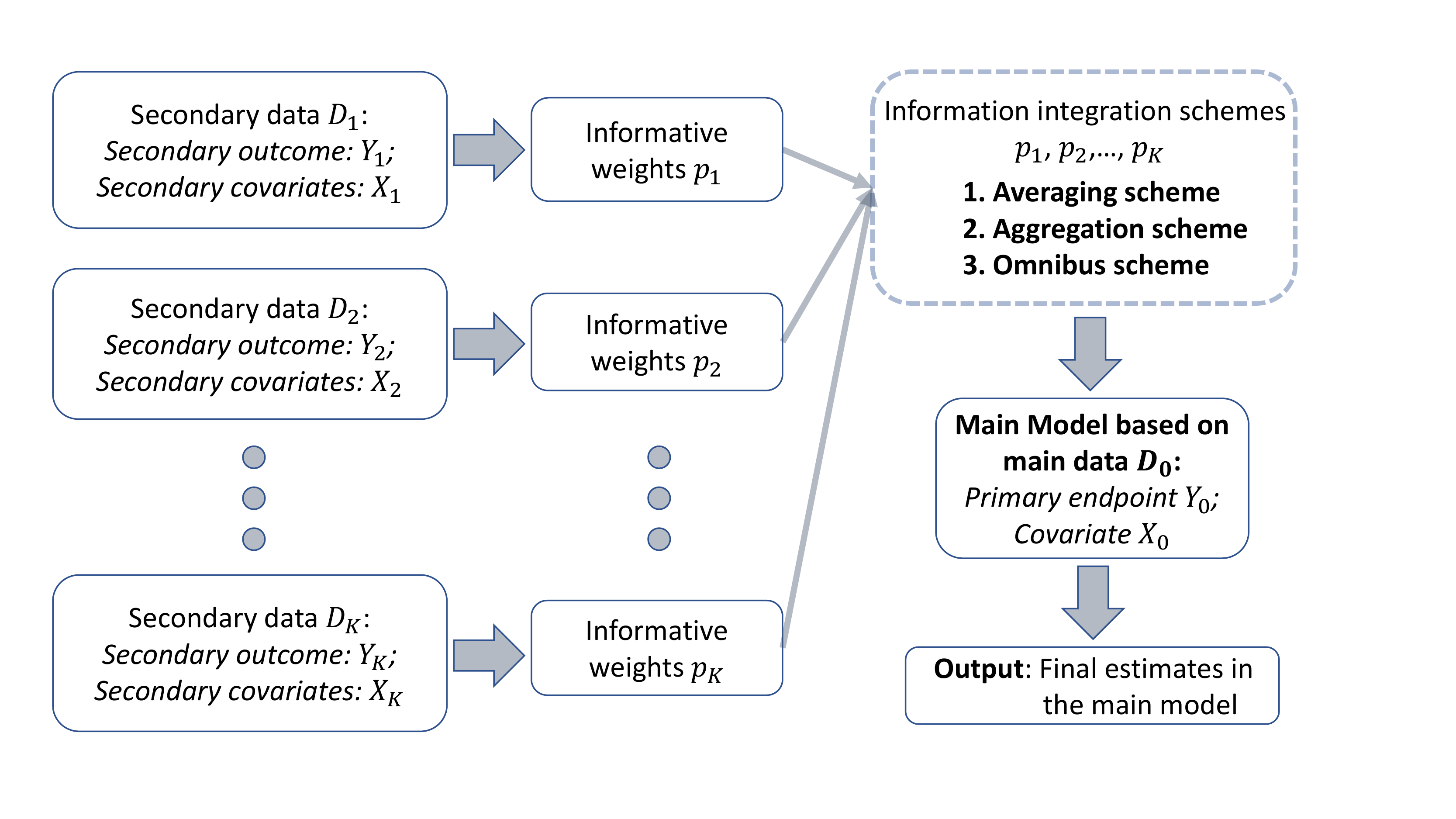}
    \caption{A schematic workflow of MinBo. MinBo first calculates informative weights based on each secondary data and then merges those weights based on several integration schemes to generate an integrated/unified score. Finally, the new score serves as a new weight in the main estimating equation to obtain parameter estimators in the main model.}\label{workflow}
\end{figure}

\revision{Figure~\ref{workflow}} is a schematic summary of the MinBo workflow. \revision{In brief, MinBo applies the empirical likelihood framework to calculate the weights for each subject. These weights are then integrated into a unified score, which is used for the main parameter estimation by re-weighting the main estimating equation. 

Before introducing the method, we first} describe the basic notation. For subject $i$ from $1,\dots, n$, let $\bfD_{0i}$ be the data collected by the main study. This dataset contains the primary endpoint $\bfY_{0i}$ and risk factors $\bfX_{0i}$ as covariates. \revisiontwo{In addition to} the primary endpoint $Y_{0i}$, there are \revisionthree{$K\geq 1$} secondary outcomes recorded for the same subject. These are highly correlated with the primary endpoint. For $k=1,\ldots,K$, let $\bfD_{ki}$ be the $k$th secondary dataset, which contains secondary outcomes $\bfY_{ki}$ with various covariates $\bfX_{ki}$. The secondary outcomes $\bfY_{ki}$ may differ from the primary endpoint in terms of distribution and data structure. \revisiontwo{Covariates $\bfX_{ki}$ in the secondary data can be the same as, partially overlapping with, or totally differently from the covariates $\bfX_{0i}$ in the main dataset.} Suppose that the model of main interest is a regression of $\bfY_{0i}$ on $\bfX_{0i}$ from the main dataset $\bfD_{0i}$. Conventionally, the parameter of interest $\bfbeta$ in regression can be solved by the following generic estimation equation:
\begin{equation}\label{classic estimating equations}
    \sum_{i=1}^n\bfg(\bfD_{0i};\bfbeta)=\bfzero,
\end{equation}
where the function $\bfg(\bfD_{0i};\bfbeta)$ could be the derivative of the least squares loss, the score function from the maximum likelihood approach or the generalized estimating equation approach, etc. Given some parameter values $\bfbeta_0$ such that $E\big\{\bfg(\bfD_{0i};\bfbeta_0)\big\}=\bfzero$, the asymptotic standard error can be derived based on the theory of the method of moments \citep{newey1994large}. For illustration, we assume that the main data are fully observed and that the main model is correctly specified. The more complicated problem of missing data in the main dataset is discussed in Section~\ref{conclusion} and Section~F in the Supporting Information. However, the estimation procedure in Equation~(\ref{classic estimating equations}) may not be efficient as it does not use any secondary data. To take into account the strong associations among the primary endpoint and secondary outcomes, \cite{chen2021improving} proposed a computationally efficient estimation approach that incorporates information from one secondary outcome into the main model. Specifically, 
the enhanced estimator can be obtained by the following re-weighted estimating equation:
\begin{equation} \label{weighted estimating equations}
    \sum_{i=1}^{n}\hat{p}_{ki}\bfg(\bfD_{0i};\bfbeta)=\bfzero,
\end{equation}
\revision{where the non-negative weights $\hat{p}_{ki}$ are estimated by maximizing} $\prod_{i=1}^np_{ki}$ \revision{with respect to $p_{ki}$ and $\bftheta_k$} under the following constraints:
\begin{equation} \label{constraints}
    \sum_{i=1}^n p_{ki}=1, \quad \sum_{i=1}^np_{ki}R_{ki}\bfh_k(\bfD_{ki};\bftheta_k)=\bfzero.
\end{equation}
The above constrained optimization problem can be solved by adopting Lagrange multipliers \citep{qin1994empirical,owen2001empirical}. $R_{ki}$ is the indicator for observing the $k$th secondary dataset for subject $i$. \revision{The estimating function $\bfh_k(\bfD_{ki};\bftheta_k)$ is derived from a working model parameterized by $\bftheta_k$ for the data $\bfD_{ki}$, such as generalized estimating equations \citep{liang1986longitudinal}}. To improve the estimation efficiency for the main parameter $\bfbeta$, the dimension of the function $\bfh_k(\bfD_{ki};\bftheta_k)$ should be larger than the dimension of the corresponding nuisance parameter vector $\bftheta_{k}$, thus rendering an over-identified function \citep{chen2021improving}. \revision{Specific examples of over-identified functions in the literature are provided in Section~G of the Supporting Information, which considers longitudinal or cross-sectional secondary outcomes.} 

Note that the estimating equations~(\ref{weighted estimating equations}) and~(\ref{constraints}) are slightly different from those proposed by \cite{chen2021improving}. \revisiontwo{In Sections~D and~H of the Supporting Information, we show that (a) these two estimation procedures are equivalent and that (b) the estimator solved by~(\ref{weighted estimating equations}) is asymptotically equivalent to a semi-parametric efficient estimator solved by empirical likelihood \citep{qin1994empirical}, given that the main function $\bfg$ and the over-identified function $\bfh$ are known.}
\revision{Further, the above method is feasible only if there is one secondary dataset. This highlights the limitation of the method in Equation~(\ref{weighted estimating equations}) for practical use when multiple secondary datasets are available. To fill this gap, we developed an efficient data integration framework that improves the estimation efficiency in the main model \revision{(Figure~\ref{workflow})}. In the remaining sections, we illustrate the framework with three integration schemes and compare their performance.}

\subsection{Averaging scheme in MinBo}\label{The averaging approach} 

{The first scheme is the so-called averaging scheme.} Specifically, we first apply~(\ref{weighted estimating equations}) and~(\ref{constraints}) to each secondary dataset. Then, based on the calculated weights $\hat{p}_{ki}$ from the $k$th working model, we construct a new score and derive the averaging estimator by solving the following re-weighted estimating equation:
\begin{equation}\label{averaging}
    \sum_{i=1}^n\bar{p}_i\bfg(\bfD_{0i};\bfbeta)=\bfzero,
\end{equation}
where $\bar{p}_i$ is obtained from a convex combination of $\hat{p}_{ki}$, i.e., $\bar{p}_i=\sum_{k=1}^K\omega_k\hat{p}_{ki}$ with non-negative constants $\omega_k$ such that $\sum_{k=1}^K\omega_k=1$. 

For illustration, we start with a simple case where two secondary datasets are available in the study ($K=2$). The generalization to $K>2$ can be found in Section~\ref{Generalization: the omnibus approach}. Let ${\bar{\bfbeta}}$ be the solution to~(\ref{averaging}). \revision{Note that the resulting estimator is consistent and robust to misspecification of any working model for the secondary data. Thus, it suffices to show that the true parameter value ${{\bfbeta}_0}$ defined in Section~\ref{Main and multiple auxiliary data} is the solution to~(\ref{averaging}) as $n\rightarrow\infty$.} \revision{By calculating the} Lagrange multiplier $\hat{\bflambda}_k$ and empirical likelihood estimator $\hat{\bftheta}_k$ (see Section~B of the Supporting Information), we can re-express each individual weight as $\hat{p}_{ki}=(1/n)/\{1+\hat{\bflambda}_k^TR_{ki}\bfh_k(\bfD_{ki}; \hat{\bftheta}_k)\}$, for $k=1,2$. Then, the estimating equation $\sum_{i=1}^n\bar{p}_i\bfg(\bfD_{0i};\bfbeta_0)=\bfzero$ can be rewritten as
\begin{equation}\label{averaging consistency}
\begin{split}
        &\frac{1}{n}\sum_{i=1}^n\bigg\{\frac{\omega_1}{1+\hat{\bflambda}_1^TR_{1i}\bfh_1(\bfD_{1i}; \hat{\bftheta}_1)}
        +\frac{\omega_2}{1+\hat{\bflambda}_2^TR_{2i}\bfh_2(\bfD_{2i}; \hat{\bftheta}_2)}\bigg\}\bfg(\bfD_{0i};\bfbeta_0)\\
        =&\frac{1}{n}\sum_{i=1}^n(\omega_1+\omega_2)\bfg(\bfD_{0i};\bfbeta_0)+o_p(\bfone)\stackrel{p}\longrightarrow  E\big\{\bfg(\bfD_{0i};\bfbeta_0)\big\}=\bfzero.
\end{split}
\end{equation}
\revision{Since $\hat{\bflambda}_k=O_p(\bfn^{-1/2})$ and $\hat{\bftheta}_k-\bftheta_k^\ast=O_p(\bfn^{-1/2})$ for all $k$ and since some $\bftheta_k^\ast$ satisfies $E\big\{R_k\bfh_k(\bfD_{ki};\bftheta_k^\ast)\big\}=\bfzero$ based on empirical likelihood theory} \citep{qin1994empirical,owen2001empirical,han2014multiply}, we can see that the above result holds. Based on~(\ref{averaging consistency}), $\bfbeta_0$ is the solution to~(\ref{averaging}) as $n\rightarrow\infty$. This finding not only implies the consistency of $\bar{\bfbeta}$ but also signifies the robust estimation of the proposed scheme. The specification of $\bfh_k(\bfD_{ki}; \bftheta^\ast_k)$ has little impact on the estimation consistency of the main parameter $\bfbeta$, as long as $E\big\{R_k\bfh_k(\bfD_{ki};\bftheta_k^\ast)\big\}=\bfzero$ holds for some value of $\bftheta_k^\ast$ (not necessarily the true one). 
This is a desirable property, as correctly specifying models for all secondary data is hard to achieve in practice. We note that model specification is vulnerable to various factors such as misspecified mean structures or failure to account for informative missingness in secondary data. The following theorem summarizes the asymptotic normality property of the resulting estimator.

\begin{theorem}\label{thm1}
Under the regularity conditions in Section~A in the Supporting Information, the estimator $\bar{\bfbeta}$ from averaging scheme asymptotically follows Normal distribution with variance-covariance matrix equal to
\begin{equation*}
\begin{split}
        &\text{Var}\big\{n^{\frac{1}{2}}(\bar{\bfbeta}-\bfbeta_0)\big\}
        =\bfvarGamma^{-1}\big\{\bfvarSigma-(2\omega_1-\omega_1^2)\bfvarLambda_1\bfS_1\bfvarLambda_1^T -(2\omega_2-\omega_2^2)\bfvarLambda_2\bfS_2\bfvarLambda_2^T\\
        &+\omega_1\omega_{2}\bfvarLambda_1\bfS_1\bfS_{h_1h_{2}}\bfS_{2}^T\bfvarLambda_{2}^T+\omega_2\omega_{1}\bfvarLambda_2\bfS_2\bfS_{h_2h_{1}}\bfS_{1}^T\bfvarLambda_{1}^T\big\}(\bfvarGamma^T)^{-1},
\end{split}
\end{equation*}
where $\bfvarGamma=E\{\partial \bfg(\bfD_{0i};\bfbeta_0)/ \partial\bfbeta^T\}$,$\bfvarLambda_k=E\{\bfg(\bfD_{0i};\bfbeta_0)\tilde{\bfh}_k(\bfD_{ki};\bftheta_k^\ast)^T\}$, $\bfvarSigma=E\{\bfg^{\otimes 2} (\bfD_{0i};\bfbeta_0)\}$, 
 and $\bfS_k=\bfS_{k11}^{-1}(\bftheta^\ast_k) -\bfS_{k11}^{-1}(\bftheta^\ast_k)\bfS_{k12}(\bftheta^\ast_k) \bfvarOmega_k(\bftheta_k^\ast) \bfS_{k21}(\bftheta_k^\ast) \bfS_{k11}^{-1}(\bftheta_k^\ast)$, for $k=1,2$. Here, $\tilde{\bfh}_k(\bfD_{ki};\bftheta_k^\ast)=R_{ki}\bfh_k(\bfD_{ki};\bftheta_k^\ast)$, $\bfvarOmega_k(\bftheta_k^\ast)=\{\bfS_{k21}(\bftheta_k^\ast)\bfS_{k11}^{-1}(\bftheta_k^\ast) \bfS_{k12}(\bftheta_k^\ast)\}^{-1}$, $\bfS_{k11}(\bftheta_k^\ast)=E\{\tilde{\bfh}_k^{\otimes 2} (\bfD_{ki};\bftheta^\ast_k)\}$, $\bfS_{k12}(\bftheta^\ast_k)=E\{\partial \tilde{\bfh}_k(\bfD_{ki};\bftheta^\ast_k)/\partial\bftheta_k^T\}$,  $\bfS_{k21}(\bftheta^\ast_k)=\bfS_{k12}^T(\bftheta^\ast_k)$, $\bfS_{h_kh_{k^\prime}}=E(\tilde{\bfh}_k(\bfD_{ki};\bftheta_k^\ast)\tilde{\bfh}^T_{k^\prime}(\bfD_{k^\prime i};\bftheta^\ast_{k^\prime}))$, for $k\neq k^\prime$. The notation $\bff^{\otimes 2} $ is $\bff\bff^T$ for any function vector $\bff$. 
\end{theorem}
\begin{remark} \label{rmk1}
It can easily be shown that the resulting asymptotic variance is no larger than the variance $\tilde{\bfV}=\bfvarGamma^{-1}\bfvarSigma(\bfvarGamma^T)^{-1}$ from the estimator solved by the unweighted estimation equation~(\ref{classic estimating equations}). 
\end{remark}

In the brackets of the covariance expression in Theorem~\ref{thm1}, the first component $\bfvarSigma$ is from the main dataset $\bfD_{0i}$. The second and third components, i.e., $\bfvarLambda_1\bfS_1\bfvarLambda_1^T$ and $\bfvarLambda_2\bfS_2\bfvarLambda_2^T$, refer to the variance reduction from integrating secondary data $\bfD_{1i}$ and $\bfD_{2i}$. The remaining two terms are due to an association between two secondary outcomes. Accordingly, the association strength among secondary outcomes can affect information borrowing. \revision{In what follows, we consider two extreme situations.}

\begin{remark} \label{rmk2}
 When two secondary datasets $\bfD_{1i}$ and $\bfD_{2i}$ are identical, i.e., $\bfD_{1i}=\bfD_{2i}=\bfD_i$, the resulting variance matrix from Theorem~\ref{thm1} reduces to the matrix found by incorporating either one of the two secondary datasets. Thus, the information is fully preserved. An intuitive explanation is that the reconstructed score is able to recover a latent and dominated quantity shared by the highly associated secondary outcomes, thus preserving the most amount of information from these highly associated secondary outcomes. 
\end{remark}

\begin{remark} \label{rmk3}
Suppose two secondary outcomes in the secondary data are independent. The resulting variance matrix reduces to $\bfvarGamma^{-1}\big\{\bfvarSigma-(2\omega_1-\omega_1^2)\bfvarLambda_1\bfS_1\bfvarLambda_1^T-(2\omega_2-\omega_2^2)\bfvarLambda_2\bfS_2\bfvarLambda_2^T\big\}(\bfvarGamma^T)^{-1}$. If the two constants $\omega_1$ and $\omega_2$ are both positive, the averaging scheme cannot aggregate the entire information from the two secondary datasets, i.e., $2\omega_1-\omega_1^2=2\omega_2-\omega_2^2=1$.
\end{remark}

Finally, the averaging scheme in Equation~(\ref{averaging}) relies on the choice of prior weights $\omega_k$. The desired weights should be able to balance the contributions from secondary outcomes. A greater weight should be assigned to a secondary dataset contributing more to the main parameter estimation. Thus, we consider weights by adopting the so-called index of information borrowing (IIB):
\begin{equation}\label{weights}
\begin{split}
    \omega_k=\frac{\text{IIB}_k}{\sum_{k^\prime=1}^K \text{IIB}_{k^\prime}}, 
    ~with~ \text{IIB}_k=\text{trace}\{\text{Diag}(\hat{\tilde{\bfV}})^{-\frac{1}{2}}\hat{\bfvarGamma}^{-1}\hat{\bfvarLambda}_k\hat{\bfS}_k\hat{\bfvarLambda}_k^T(\hat{\bfvarGamma}^T)^{-1}\text{Diag}(\hat{\tilde{\bfV}})^{-\frac{1}{2}}\},
\end{split}
\end{equation}
where the terms $\hat{\tilde{\bfV}}$, $\hat{\bfvarGamma}$, $\hat{\bfvarLambda}_k$, and $\hat{\bfS}_k$ are consistent estimators of $\tilde{\bfV}$, $\bfvarGamma$, $\bfvarLambda_k$, and $\bfS_k$, respectively, \revision{and $\text{Diag}(\hat{\tilde{\bfV}})$ denotes a matrix with diagonal elements equal to $\hat{\tilde{\bfV}}$ and other elements equal to zeros.} 
Note that the term $\bfvarGamma^{-1}\bfvarLambda_k\bfS_k\bfvarLambda_k^T(\bfvarGamma^T)^{-1}$ is the reduced estimation variability due to incorporating the $k$th secondary dataset. The term $\text{Diag}(\tilde{\bfV})^{-1/2}$ makes the efficiency gain comparable with that of the unweighted estimator in Equation~(\ref{classic estimating equations}). As $\bfS_k$ is non-negative definite, the defined weights are always non-negative. A larger value indicates better performance in delivering information from the $k$th secondary dataset to the main analysis. Thus, the proposed weights in~(\ref{weights}) will assign more credit to secondary datasets with a higher contribution to the main analysis.

\subsection{Aggregating scheme in MinBo}
\label{The aggregating approach}
\revision{The second scheme in MinBo is called the aggregating scheme.} Suppose the individual weights $\hat{p}_{ki}$ for the $k$th dataset, with $i=1,\ldots,n$, are calculated as described in Section~\ref{Main and multiple auxiliary data}. Then, we consider the following re-weighting scheme:
\begin{equation}\label{aggregating}
    \sum_{i=1}^n\tilde{p}_i\bfg(\bfD_{0i};\bfbeta)=\bfzero,
\end{equation}
where the aggregated scores $\tilde{p}_i$ are defined as $\tilde{p}_i=\prod_{k=1}^K\hat{p}_{ki}$. Again for illustration, we focus on the situation with two secondary datasets ($K=2$). 
We denote the solution to~(\ref{aggregating}) by $\tilde{\bfbeta}$. \revision{As in Section~\ref{The averaging approach}, we can show that the true ${\bfbeta}_0$ is the solution to~(\ref{aggregating}) as $n\rightarrow\infty$ as}
\begin{equation}\label{aggregating consistency}
    \begin{split}
        &\frac{1}{n}\sum_{i=1}^n\bigg\{\frac{1}{1+\hat{\bflambda}_1^TR_{1i}\bfh_1(\bfD_{1i}; \hat{\bftheta}_1)}\bigg\}
        \cdot\bigg\{\frac{1}{1+\hat{\bflambda}_2^TR_{2i}\bfh_2(\bfD_{2i}; \hat{\bftheta}_2)}\bigg\}\bfg(\bfD_{0i};\bfbeta_0)\\
        =&\frac{1}{n}\sum_{i=1}^n\bfg(\bfD_{0i};\bfbeta_0)+o_p(\bfone)
        \stackrel{p}\longrightarrow E\big\{\bfg(\bfD_{0i};\bfbeta_0)\big\}=\bfzero.
\end{split}
\end{equation}
Accordingly, the derivation implies that the estimator $\tilde{\bfbeta}$ is a consistent estimator and robust to a misspecification of the working model for secondary data \revision{if $E\big\{R_k\bfh_k(\bfD_{ki};\bftheta_k^\ast)\big\}=\bfzero$.} The following theorem summarizes the asymptotic property of the estimator via aggregation.

\begin{theorem}\label{thm2}
Under the regularity conditions in Section~A in the Supporting Information, the estimator $\tilde{\bfbeta}$ from aggregating scheme asymptotically follows Normal distribution with variance-covariance matrix equal to
\begin{equation*}
\begin{split}
        \text{Var}&\big\{n^{\frac{1}{2}}(\tilde{\bfbeta}-\bfbeta_0)\big\}=\bfvarGamma^{-1}\big\{\bfvarSigma-\bfvarLambda_1\bfS_1\bfvarLambda_1^T-\bfvarLambda_2\bfS_2\bfvarLambda_2^T\\
        &+\bfvarLambda_1\bfS_1\bfS_{h_1h_2}\bfS_2^T\bfvarLambda_2^T+\bfvarLambda_2\bfS_2\bfS_{h_2h_1}\bfS_1^T\bfvarLambda_1^T\big\}(\bfvarGamma^T)^{-1}.
\end{split}
\end{equation*}
All notations are defined in Theorem~\ref{thm1}. 
\end{theorem}

\begin{remark}\label{rmk4}
\revision{ Let us revisit the situation in Remark~\ref{rmk3}.} Suppose two secondary outcomes from secondary datasets $\bfD_{1i}$ and $\bfD_{2i}$ are independent. The resulting asymptotic variance reduces to $\bfvarGamma^{-1}\big\{\bfvarSigma-\bfvarLambda_1\bfS_1\bfvarLambda_1^T-\bfvarLambda_2\bfS_2\bfvarLambda_2^T\big\}(\bfvarGamma^T)^{-1}$. This implies that the aggregation scheme is able to recover all the information from the two independent secondary datasets. Thus, the aggregating scheme is better than the averaging scheme if there is only a mild association among secondary datasets/outcomes.  
\end{remark}

\begin{remark} \label{rmk5}
Suppose two secondary datasets are identical, i.e., $\bfD_{1i}=\bfD_{2i}=\bfD_{i}$. The resulting asymptotic variance reduces to $\tilde{\bfV}$, i.e., the variance from the unweighted estimation procedure in Equation~(\ref{classic estimating equations}). Thus, there is no efficiency gain from aggregating two such highly associated secondary datasets/outcomes.
\end{remark}

\subsection{Generalization of MinBo: The omnibus approach}\label{Generalization: the omnibus approach}
 We have described two competing schemes for integrating information from multiple secondary datasets. In summary, the averaging scheme is preferable when the secondary outcomes are highly associated (Remark~\ref{rmk2}), whereas the aggregating approach is preferable when the secondary outcomes are mildly associated (Remark~\ref{rmk4}). Now, we generalize MinBo by introducing a third scheme, the omnibus scheme, which is more versatile and includes desirable properties from the other two schemes. \revision{We first present a general setting and then illustrate three practical cases at the end of this section.}

For $k=1,\ldots,K$ and $k^\prime=1,\ldots,K^\prime$ with some integer $K^\prime\leq K$, let us define a user-specified array containing elements with non-negative values $\omega_{k^\prime k}$ such that for every $k^\prime$, we have $\sum_{k=1}^K\omega_{k^\prime k}=1$. Suppose the individual weights $\hat{p}_{ki}$ are already available based on the procedure in Equation~(\ref{constraints}) corresponding to each secondary dataset. Then, the omnibus estimator is derived by solving the following re-weighted estimating equation:
\begin{equation}\label{omnibus}
    \sum_{i=1}^n\check{p}_i\bfg(\bfD_{0i};\bfbeta)=\bfzero,
\end{equation}
where the omnibus scores are calculated as
\begin{equation}\label{combined propensity scores}
    \check{p}_i=\prod_{k^\prime=1}^{K^\prime}\bigg(\sum_{k=1}^K\omega_{k^\prime k}\hat{p}_{k i}\bigg).
\end{equation}

\begin{remark}\label{rmk6}
The omnibus estimation scheme is a combination of the averaging and aggregating schemes. Accordingly, from the arguments in Sections~\ref{The averaging approach} and~\ref{The aggregating approach}, we can state that the resulting omnibus estimator $\check{\bfbeta}$ is also consistent and robust to a misspecification of the working model for any secondary data under a mild first-moment condition. 
\end{remark}

\begin{theorem}\label{thm3}
Under the regularity conditions in Section~A in Supporting Information, the resulting estimator $\check{\bfbeta}$ asymptotically follows Normal distribution with variance-covariance matrix equal to
\begin{equation*}
\begin{split}
        &\text{Var}\big(n^{\frac{1}{2}}(\check{\bfbeta}-\bfbeta_0)\big)=\bfvarGamma^{-1}\bigg\{\bfvarSigma-\sum_{m=1}^K(2\tilde{\omega}_m-\tilde{\omega}_m^2)\bfvarLambda_m\bfS_m\bfvarLambda_m^T\\
        &+\sum_{m\neq m^\prime}^K\tilde{\omega}_{m}\tilde{\omega}_{m^\prime}\bfvarLambda_m\bfS_m\bfS_{h_mh_{m^\prime}}\bfS_{m^\prime}^T\bfvarLambda_{m^\prime}^T\bigg\}(\bfvarGamma^T)^{-1},
\end{split}
\end{equation*}
where $\tilde{\omega}_m=\sum_{k^\prime=1}^{K^\prime}\omega_{k^\prime m}$. All notations, in particular for $\bfvarLambda_m$, $\bfS_m$, and $\bfS_{h_mh_{m^\prime}}$ would be the same to Theorem~\ref{thm1}, except that now the index $m$ and $m^\prime$ could be larger than two.
\end{theorem}
 
  When $K^\prime=1$, Equation~(\ref{omnibus}) reduces to the averaging scheme in Equation~(\ref{averaging}). When $\omega_{k^\prime k}=1$ for $k^\prime=k$ and $\omega_{k^\prime k}=0$ otherwise, Equation~(\ref{omnibus}) becomes the aggregating scheme in Equation~(\ref{aggregating}). Moreover, the omnibus scheme is more versatile, as it can handle more complex situations in practice.
  
Let us consider three common situations. Suppose we have three secondary datasets $\bfD_{1i}$, $\bfD_{2i}$, and $\bfD_{3i}$, in addition to the main study dataset $\bfD_{0i}$. If the secondary outcomes are slightly correlated, we apply the aggregating scheme by setting $\omega_{11}=\omega_{22}=\omega_{33}=1$ and $\omega_{k^\prime k}=0$ for $k^\prime\neq k$. In contrast, if all the secondary outcomes are highly correlated, we apply the averaging scheme by setting $K^\prime=1$. If the secondary outcomes from the first two datasets are highly correlated but only slightly correlated with the outcome in the third secondary dataset, we \revision{can} apply a combination such that $\check{p}_i=(\omega_{11}\hat{p}_{1i}+\omega_{12}\hat{p}_{2i})\hat{p}_{3i}$, where we set $K^\prime=2$, $\omega_{23}=1$, and $\omega_{13}=\omega_{21}=\omega_{22}=0$. Given such a reconstructed score, the problem due to the high association between $\bfD_{1i}$ and $\bfD_{2i}$ can be alleviated by a convex combination of $\hat{p}_{1i}$ and $\hat{p}_{2i}$. Thus, information from  $\bfD_{3i}$ can be substantially aggregated into the main analysis. The weights for these three practical situations can be determined with~(\ref{weights}) in Section~\ref{The averaging approach}. \revision{For practical use, the form of the omnibus approach can initially be determined from the clinical features of the secondary outcomes in terms of their association and then refined by comparing it with one or two of the competing forms based on the estimated efficiency gain.} 

\section{Simulations}\label{Simulation Studies}
In this section, we evaluated the numerical performance of our proposed estimators in MinBo. To mimic the situation in the ARIC study but also to keep the setup in a general context, we considered the following setting \revision{(Figure~S1 of the Supporting Information)}. There was one main dataset $\bfD_{0i}$ consisting of a primary endpoint in a binary scale. There were also three secondary datasets, $\bfD_{1i}$, $\bfD_{2i}$, and $\bfD_{3i}$, collected from the same study. The first two secondary datasets $\bfD_{1i}$ and $\bfD_{2i}$ contained repeated measurements on a continuous scale. The third secondary dataset $\bfD_{3i}$ contained a binary variable in cross-sectional format.
In the following, we generated different association patterns among the three secondary outcomes to assess three proposed data integration schemes and check the robustness of MinBo when all working models for the three secondary datasets were subject to misspecification.

\subsection{Data generation}\label{Data generation under correct secondary models}
First, we generated the three secondary datasets. The outcomes in the first and second secondary datasets were longitudinal measurements modeled by $\bfY_{ki}=\bfX_{ki}\bftheta_k+\bfepsilon_{ki}$ for $k=1,2$ with $\bftheta_1=(-1,2,1,1)^T$, $\bftheta_2=(1,-2,-1,-1)^T$, and $\bfX_{ki}=(\bfone,\tilde{\bfX}_{ki1},\tilde{\bfX}_{ki2},\tilde{\bfX}_{ki3})$ where $\tilde{\bfX}_{kij}=(\tilde{X}_{kij1},\ldots,\tilde{X}_{kij4})^T$. Detailed specifications of $\bfX_{ki}$ and $\bfepsilon_{ki}$ are given in Section~E of the Supporting Information. We used $\rho$ to quantify the correlation between $\bfepsilon_{1i}$ and $\bfepsilon_{2i}$ with three different values $\rho=0$, 0.4, or 0.8 representing no, mild, and strong correlations, respectively. Moreover, in the third dataset $\bfD_{3i}$, the outcome $Y_{3i}$ followed a Bernoulli distribution with success probability $p_i^\ast=\{1+\exp(-{\bfX}_{3i}^T\bftheta_3)\}^{-1}$, where $\bftheta_3=(-1,1)^T$ and ${\bfX}_{3i}=(1,\tilde{X}_{1i11})^T$ with $X_{1i11}$ defined in the first dataset $\bfD_{1i}$. 

To build possible associations between outcome $Y_{3i}$ and outcomes $\bfY_{1i}$ and $\bfY_{2i}$, we considered the following generation mechanism. For each $i$, the binary outcome $Y_{3i}$ was equal to $1$ if $\epsilon_{1i1}\geq x_{i}^\ast$ and equal to $0$ otherwise, where $\epsilon_{1i1}$ was the first element in residual  vector $\bfepsilon_{1i}$ and $x_{i}^\ast$ was the $(1-p_i^\ast)$th percentile of the standard normal distribution. 
Thus, the resulting outcome $Y_{3i}$ was always associated with repeated measurements $\bfY_{1i}$, whereas it was independent of the outcome $\bfY_{2i}$ if $\rho=0$ but highly associated with $\bfY_{2i}$ if $\rho=0.8$ \revision{(Figure~S1 of the Supporting Information)}. 

We next generated the main dataset $\bfD_{0i}$. The outcome of primary interest $Y_{0i}$ in the main dataset $\bfD_{0i}$ was considered to be binary with success probability $p_{0i}=\{1+\exp(-{\bfX}_{0i}^T\bfbeta_0)\}^{-1}$ with $\bfbeta_0=(1,-0.5,-1,0.5)^T$ and ${\bfX}_{0i}=(1,\tilde{X}_{1i11},\tilde{X}_{1i21},\tilde{X}_{1i31})^T$. To ensure it was associated with the three secondary outcomes, we considered the following generation procedure. For each $i$, we constructed a variable $\bar{x}_i=(\epsilon_{1i4}+\epsilon_{2i4})/\alpha$ with the scalar $\alpha$ specified to make $\bar{x}_i$ follow a standard normal distribution. Then, the primary endpoint $Y_{0i}$ was equal to $1$ if $\bar{x}_i\geq x_{0i}$ and equal to $0$ otherwise. The value $x_{0i}$ was the $(1-p_{0i})$th percentile of the standard normal distribution. Thus, the primary endpoint $Y_{0i}$ was associated with all three secondary outcomes.
\revision{We also considered and evaluate the case where the secondary outcomes were \revisiontwo{slightly} associated with the primary endpoint. (The details are in Section~E of the Supporting Information.)} 

\subsection{Evaluating the method}\label{sec:3.2}

Based on the main dataset $\bfD_{0i}$, the unweighted estimation equation~(\ref{classic estimating equations}) was the score function from a logistic regression. The weights $\hat{p}_{1i}$ and $\hat{p}_{2i}$ for the first two secondary datasets $\bfD_{1i}$ and $\bfD_{2i}$ were calculated via~(\ref{constraints}),  where working functions $\bfh_1(\bfD_{1i};\bftheta_1)$ and $\bfh_2(\bfD_{2i};\bftheta_2)$ were based on the specification in Section~G of the Supporting Information. 
The weights $\hat{p}_{3i}$ were calculated based on the working function $\bfh_3(\bfD_{3i};\bftheta_3)$ with $\tilde{Z}_{3i}=(\tilde{X}_{1i21},\tilde{X}_{1i31})^T$ as a redundant variable in Section~G of the Supporting Information. We considered the combination of sample sizes $n=300$ or $600$ with $\rho=0$, $0.4$, or $0.8$. 
Both fully and partially observed secondary data were considered in our setup. The three proportions of observing each secondary dataset were set to $\eta_1=0.6$, $\eta_2=0.7$, and $\eta_3=0.5$, respectively. 

Thereafter, we compared eight estimators in total under different setups. In particular, estimators \text{single100}, \text{single010}, and  \text{single001}  incorporated the first, second, or third secondary dataset, respectively. Estimators \text{ave110} and \text{agg110} incorporated $\bfD_{1i}$ and $\bfD_{2i}$ via the averaging scheme or the aggregating scheme, respectively. Estimators \text{ave111} and  \text{agg111} incorporated the three secondary datasets via the averaging scheme or the aggregating scheme, respectively. Finally, estimator \text{omn111} incorporated the three secondary datasets via the omnibus scheme.  The estimator \text{omn111} used the combined score defined as $\check{p}_i=(\omega_{11}\hat{p}_{1i}+\omega_{13}\hat{p}_{3i})\hat{p}_{2i}$. All the weights in the averaging and omnibus schemes were specified based on~(\ref{weights}).

 Table~\ref{table1} summarizes the estimation bias, Monte Carlo standard deviation, asymptotic standard error, and $95\%$ coverage probability of the eight estimators based on $1000$ Monte Carlo runs, given $\rho=0.8$, i.e., a high correlation among the three secondary outcomes. For the small and large sample sizes, we observed that all estimators had a small bias and a satisfactory coverage probability close to $95\%$. As the sample size increased from $300$ to $600$, the bias became closer to zero. We also observed that the standard errors for fully observed secondary data were smaller than those for partially observed secondary data. Results with $\rho=0.4$ and $\rho=0$ are displayed in Tables~S1 and~S2 of the Supporting Information, respectively. The same patterns were observed in these tables.

We used the empirical relative efficiency (ERE) to assess the efficiency gain for the eight competing estimators compared with the MLE method. ERE is the ratio between the empirical variance of the MLE via logistic regression without considering any secondary data and the empirical variance of the proposed estimator. Larger values (larger than $1$) are better. Table~\ref{table2} summarizes the results with fully observed secondary data.

First, we considered estimators based on the existing method incorporating only one secondary dataset (\text{single100}, \text{single010}, and \text{single001}). ERE for $\beta_{3}$ was barely above 1 when incorporating either $\bfD_{1i}$ or \revision{$\bfD_{3i}$}. ERE for $\beta_{1}$ was low when integrating \revision{$\bfD_{2i}$}, which \revision{implies that the incorporation of only one secondary dataset makes only a limited contribution to the main estimation.} Next, we evaluated how the associations among secondary outcomes affected the data integration by comparing \text{ave110} and \text{agg110}.
When $\rho=0$, i.e., there was no association between $\bfY_{1i}$ and $\bfY_{2i}$, \text{agg110} slightly outperformed \text{ave110} in terms of ERE. {However, \text{agg110} performed worse as $\rho$ became closer to 1}, whereas \text{ave110} performed increasingly better as the correlation coefficient increased. These results are consistent with the theoretical properties (Remarks~\ref{rmk2},~\ref{rmk3},~\ref{rmk4}, and~\ref{rmk5}). {Note that \text{ave110} also had a better ERE than \text{single100} and \text{single010} for all values of $\rho$. 
These findings justify the conclusion in Sections~\ref{The averaging approach} and~\ref{The aggregating approach} that the aggregating scheme leads to a better fit when there is a mild correlation between secondary outcomes, whereas the averaging scheme is preferable given strongly correlated secondary outcomes.} Finally, we evaluated MinBo by integrating the three secondary datasets. In this case, \text{omn111} with $\check{p}_i=(\omega_{11}\hat{p}_{1i}+\omega_{31}\hat{p}_{3i})\hat{p}_{2i}$ performed the best in terms of ERE when $\rho=0$ or $0.4$. The estimator \text{ave111} was the best when $\rho=0.8$.

Note that for all setups, the best estimator among \text{ave111}, \text{agg111}, and \text{omn111} was also better than all estimators incorporating only one or two secondary datasets, in terms of having a higher ERE. Thus, correctly integrating multiple secondary datasets facilitates a gain in estimation efficiency in the main model. These findings also applied to cases where partially observed secondary data were available (Table~S3).

\subsection{Misspecified working models}\label{Misspecified secondary models}\label{sec:3.3}

As stated in Remark~\ref{rmk6}, misspecification of the working model for any secondary data may have little impact on the estimation consistency of the main model in MinBo. Such model misspecification in practical situations could include a failure to correctly specify the mean structures or properly account for complex missingness caused by design issues, participant refusal, patient dropouts, etc. 

\revision{To check the robustness of MinBo, we implemented a case study in which the mean structures were misspecified in all the working models when fitting three secondary datasets. The method for generating data was the same as in 
Section~\ref{Data generation under correct secondary models}}. Specifically, for secondary outcomes $\bfY_{1i}$ and $\bfY_{2i}$, rather than utilizing $\bfX_{ki}=(\bfone,\tilde{\bfX}_{ki1},\tilde{\bfX}_{ki2},\tilde{\bfX}_{ki3})$ to fit the data, we instead used a misspecified covariate vector $\bfX^{\text{mis}}_{ki}=(\bfone,\tilde{\bfX}_{ki1},\tilde{\bfX}_{ki2})$ for $k=1,2$. 
When fitting the data for secondary outcome $Y_{3i}$, we replaced the true values $\bfX_{3i}=(1,\tilde{X}_{1i11})^T$ with the covariate vector $\bfX^{\text{mis}}_{3i}=(1,\tilde{X}_{1i12})^T$. Thus, the redundant covariate $\tilde{Z}_{3i}$ contained only $\tilde{X}_{1i31}$. 

The results for fully observed secondary data are summarized in Table~\ref{table3}. With misspecified mean structures, the proposed estimators in the main model still showed little estimation bias, with a satisfactory coverage probability close to the $95\%$ nominal level. In terms of ERE, MinBo was still able to improve the estimation efficiency. Similar patterns were observed for partially observed secondary data (Table~S4). 

The above results assumed that the secondary data were complete or partially observed due to data being missing completely at random. We also investigated the impact of informative missingness \citep{enders2010applied} in the secondary data on the main parameter estimation. The data-generating scheme and results are summarized in Table~S5 of the Supporting Information. \revisiontwo{Furthermore, We evaluated the performance of MinBo under various setups, including scenarios where the secondary outcomes were \revisiontwo{mildly} associated with the main outcome (Table~S6) and a setup with a higher correlation between the main outcome and secondary outcomes (Table~S7). We also compared MinBo with the joint likelihood estimate when the joint likelihood of multiple outcomes can be specified (Table~S8).}  All the results demonstrated the robustness and satisfactory performance of MinBo. There was little bias and improved estimation efficiency.

\section{A real data application}\label{An real data application}

In this section, we applied MinBo to identify baseline risk factors for the development of essential hypertension in the white male population from center B of the ARIC study. Essential hypertension was defined as having at least one of three events: SBP $\geq 140$~mm~Hg, DBP $\geq 90$~mm~Hg, or taking anti-hypertensive medications within a 2-week period. To focus on incident cases of hypertension, we excluded subjects who had hypertension at the baseline. The primary endpoint was defined as the occurrence of hypertension (binary and cross-sectional) during the follow-up. We considered the following potential baseline risk factors: body mass index (kg/m$^2$), current alcohol drinking status (1=Yes, 0=No), current cigarette smoking status (1=Yes, 0=No), age (years), and hemoglobin (g/dL). The main dataset $\bfD_{0i}$ included 1138 subjects. The conventional estimates of risk-factor effects were derived from logistic regression (using MLE). To obtain unweighted estimates, the score function was that in Equation~(\ref{classic estimating equations}).

We considered the following three secondary datasets, which had different secondary outcomes: SBP ($\bfD_{1i}$), DBP ($\bfD_{2i}$), and ever having taken anti-hypertensive medication during the follow-up ($\bfD_{3i}$). Note that the first two secondary outcomes were longitudinal measurements recorded over four doctor visits. We adopted the estimating function in \revision{Section~G of the Supporting Information} with covariates body mass index, current alcohol drinking status, current cigarette smoking status, age, and hemoglobin level at each visit. Four basis matrices $\bfV_1$, $\bfV_2$, $\bfV_3$, and $\bfV_4$, as described in Section~\ref{Simulation Studies}, were used to construct the equations. For all subjects in the main analysis, SBP and DBP were measured at each visit.
In contrast, for the third working model, we used the estimating function in Section~G of the Supporting Information with body mass index, age, and hemoglobin as the main covariates in $\bfX_{3i}$. Current alcohol drinking status and current cigarette smoking status were redundant covariates in $\tilde{\bfZ}_{3i}$. All covariates in $\bfD_{3i}$ were measured at the baseline. Note that the main and redundant variables in $\bfD_{3i}$ in Section~G of the Supporting Information were selected using the Akaike information criterion.
All subjects in the main analysis were included in $\bfD_{3i}$. 

We calculated and compared seven estimators. One was based on the MLE approach without incorporating any secondary data. Three were based on integrating information from a single secondary dataset \citep{chen2021improving}. These were single100, single010, and single001 and incorporated the first, second, or third secondary dataset, respectively.
The remaining three estimators were from MinBo. The estimators agg111 and ave111 integrated $\bfD_{1i}$, $\bfD_{2i}$, and $\bfD_{3i}$ via the aggregating scheme or the averaging scheme, respectively. 
The estimator omn111 integrated $\bfD_{1i}$, $\bfD_{2i}$, and $\bfD_{3i}$ via the omnibus approach with the unified score calculated as $\check{p}_i=(\omega_{11}\hat{p}_{1i}+\omega_{12}\hat{p}_{2i})\hat{p}_{3i}$. This omnibus estimator leveraged the strong associations between SBP and DBP due to their clinical features \revision{(conditional correlation $=0.66$)}. It aggregated the information on medication use \revision{due to a mild correlation with the other two covariates (conditional correlation $=0.2$)}. All weights were specified based on~(\ref{weights}) in Section~\ref{The averaging approach}.

Table~\ref{table4} compares the results for the six estimators (excluding MLE). The estimated relative efficiency was defined as the ratio of the estimated variance of MLE without secondary data to the estimated variance of the estimator based on data integration. We found that all estimators integrating secondary data performed better than the MLE estimator based solely on the main data (i.e., the relative efficiency was greater than 1).
Moreover, the estimators that integrated multiple secondary datasets (ave111, agg111, and omn111) outperformed the estimators based on existing methods that use only a single secondary dataset (single100, single010, and single001). Among the estimators that used multiple secondary datasets, agg111 and omn111 from MinBo performed best, as they had the highest efficiency gain for all parameters. The efficiency gain with agg111 and omn111 was almost threefold for drinking and smoking, and around twofold for hemoglobin, compared with the MLE method. These results thereby highlight the utility of MinBo in integrating multiple secondary datasets. We present detailed results for the agg111 and omn111 estimators in Table~\ref{table5}.
The risk factors body mass index, alcohol drinking, and age were statistically significant. These findings are consistent with previous reports in the literature \citep{ shihab2012body}. Note that only body mass index and age showed evidence of significance with the MLE approach based solely on the main data, further demonstrating the superiority of MinBo.

\section{Conclusion}
\label{conclusion}
Synthesizing information from secondary data into the main data analysis is a timely research topic in epidemiology and clinical trials, as it has become much easier to collect data due to technological advances. MinBo is an initial attempt to efficiently incorporate multiple secondary datasets into the main analysis. This approach improves estimation efficiency and the consistency of the estimator in the main data model, even if the working models for any secondary data are misspecified. \revision{Additionally, MinBo can produce separate estimates for each working model, resulting in a lighter computational load compared to a joint likelihood estimate (if applicable).} 

\revisiontwo{In practice, integrating multiple secondary datasets is challenging and dependent on various factors, such as associations among secondary outcomes.}
The omnibus integration scheme in MinBo is versatile in handling such circumstances. \revision{The specific form of the omnibus approach can be determined based on clinical features of the secondary outcomes under consideration or validated through the relative efficiency calculated from the estimated variances of estimates with and without considering secondary outcomes.} 
Throughout this work, we assumed that all the main data points were observed or missing completely at random, which may not be the case in some applications. However, our method can easily be embedded into a well-developed scheme, such as the inverse probability weight technique \citep{enders2010applied,chen2021multiple}. More details of the estimation procedure are given in Section~F of the Supporting Information. Moreover, extending our data integration scheme to other statistical fields, such as survival analysis, casual inference, and high-dimensional estimation, would be of great interest and merit future work.  

\revisiontwo{\section*{Data Availability Statement}
The data that support the findings \revisionthree{in this paper} are available from the corresponding author upon reasonable request.}

\label{lastpage}

\bibliographystyle{biom} 
\bibliography{biomsample_bib}

\revisiontwo{
\section*{Supporting Information}
Web Appendices and Tables  referenced in Sections~\ref{sec:3.2} and~\ref{sec:3.3} and the R codes implementing Sections~\ref{Simulation Studies} and~\ref{An real data application} are available with this paper from
the Biometrics website on Wiley Online Library.}

\begin{table}
\caption{Overall evaluations for eight estimators (defined in Section~\ref{Simulation Studies}) under the combination of sample sizes ($n=300,600$) and proportions for observing three secondary datasets (CASE 1: full observation $\eta_1=\eta_2=\eta_3=1$; CASE 2: partial observation $\eta_1=0.6$,$\eta_2=0.7$,$\eta_3=0.5$), given the correlation ($\rho$) between two longitudinal secondary outcomes equal to $0.8$. MCSD: Monte Carlo standard deviation; ASE: asymptotic standard error; CP: coverage probability. All values have been  multiplied by $100$.}
\label{table1}
\centering
\scriptsize
\begin{tabular}{ccccccclcccc}
\toprule
           &           &         & \multicolumn{4}{c}{n=300}  &  & \multicolumn{4}{c}{n=600}  \\ \cline{4-7} \cline{9-12} 
           &           &         & Bias & MCSD & ASE & 95\%CP &  & Bias & MCSD & ASE & 95\%CP \\\midrule
CASE 1 & single100 & $\beta_0$ & 2.5  & 25   & 23  & 94     &  & 0.6  & 17   & 16  & 95     \\
           &           & $\beta_1$ & -0.5 & 25   & 24  & 93     &  & 0.2  & 18   & 17  & 94     \\
           &           & $\beta_2$ & -2.5 & 15   & 15  & 95     &  & -1.0 & 11   & 10  & 94     \\
           &           & $\beta_3$ & -0.7 & 29   & 28  & 95     &  & 0.2  & 20   & 20  & 95     \\
           & single010 & $\beta_0$ & 2.4  & 24   & 24  & 96     &  & 0.9  & 17   & 17  & 95     \\
           &           & $\beta_1$ & -0.1 & 28   & 28  & 96     &  & -0.2 & 20   & 20  & 95     \\
           &           & $\beta_2$ & -2.3 & 15   & 15  & 95     &  & -1.0 & 11   & 10  & 94     \\
           &           & $\beta_3$ & -1.0 & 25   & 24  & 95     &  & 0.2  & 17   & 17  & 95     \\
           & single001 & $\beta_0$ & 2.5  & 25   & 23  & 95     &  & 0.7  & 17   & 16  & 95     \\
           &           & $\beta_1$ & -0.7 & 26   & 25  & 94     &  & -0.1 & 18   & 17  & 94     \\
           &           & $\beta_2$ & -2.4 & 15   & 15  & 94     &  & -0.9 & 11   & 10  & 94     \\
           &           & $\beta_3$ & -0.4 & 29   & 28  & 94     &  & 0.2  & 20   & 19  & 95     \\
           & ave110    & $\beta_0$ & 2.5  & 24   & 23  & 94     &  & 0.7  & 17   & 16  & 95     \\
           &           & $\beta_1$ & -0.6 & 25   & 24  & 94     &  & 0.1  & 18   & 17  & 94     \\
           &           & $\beta_2$ & -2.4 & 15   & 14  & 95     &  & -1.0 & 11   & 10  & 95     \\
           &           & $\beta_3$ & -0.6 & 29   & 27  & 95     &  & 0.2  & 20   & 19  & 94     \\
           & agg110    & $\beta_0$ & 2.2  & 26   & 24  & 93     &  & 0.2  & 17   & 17  & 94     \\
           &           & $\beta_1$ & -0.7 & 29   & 27  & 93     &  & 0.4  & 20   & 19  & 93     \\
           &           & $\beta_2$ & -1.8 & 17   & 16  & 92     &  & -0.7 & 12   & 11  & 94     \\
           &           & $\beta_3$ & -0.8 & 30   & 27  & 94     &  & 0.0  & 20   & 19  & 94     \\
           & ave111    & $\beta_0$ & 2.4  & 24   & 23  & 95     &  & 0.7  & 16   & 16  & 95     \\
           &           & $\beta_1$ & -0.4 & 25   & 24  & 94     &  & 0.0  & 18   & 17  & 94     \\
           &           & $\beta_2$ & -2.3 & 15   & 14  & 95     &  & -0.9 & 11   & 10  & 94     \\
           &           & $\beta_3$ & -0.7 & 27   & 26  & 95     &  & 0.2  & 18   & 18  & 94     \\
           & agg111    & $\beta_0$ & 2.4  & 25   & 23  & 93     &  & 0.2  & 17   & 16  & 94     \\
           &           & $\beta_1$ & -0.7 & 29   & 27  & 93     &  & 0.4  & 20   & 19  & 93     \\
           &           & $\beta_2$ & -1.5 & 19   & 18  & 94     &  & -0.6 & 13   & 12  & 95     \\
           &           & $\beta_3$ & -1.3 & 26   & 24  & 93     &  & 0.0  & 18   & 17  & 94     \\
           & omn111    & $\beta_0$ & 2.3  & 24   & 22  & 93     &  & 0.4  & 16   & 16  & 95     \\
           &           & $\beta_1$ & -0.7 & 26   & 25  & 93     &  & 0.2  & 19   & 17  & 94     \\
           &           & $\beta_2$ & -1.8 & 16   & 15  & 93     &  & -0.7 & 11   & 10  & 94     \\
           &           & $\beta_3$ & -0.9 & 26   & 25  & 94     &  & 0.1  & 18   & 18  & 94   \\ 
\toprule
           &           &         & \multicolumn{4}{c}{n=300}  &  & \multicolumn{4}{c}{n=600}  \\ \cline{4-7} \cline{9-12} 
           &           &         & Bias & MCSD & ASE & 95\%CP &  & Bias & MCSD & ASE & 95\%CP \\\midrule
CASE 2  & single100 & $\beta_0$ & 2.6  & 25   & 24  & 95     &  & 0.9  & 17   & 17  & 95     \\
           &           & $\beta_1$ & -0.3 & 27   & 26  & 94     &  & -0.1 & 19   & 18  & 94     \\
           &           & $\beta_2$ & -2.8 & 16   & 15  & 94     &  & -1.1 & 11   & 11  & 94     \\
           &           & $\beta_3$ & -0.7 & 29   & 28  & 95     &  & 0.3  & 20   & 20  & 95     \\
           & single010 & $\beta_0$ & 2.6  & 25   & 24  & 95     &  & 0.8  & 17   & 17  & 95     \\
           &           & $\beta_1$ & -0.1 & 28   & 28  & 96     &  & -0.2 & 20   & 20  & 95     \\
           &           & $\beta_2$ & -2.5 & 15   & 16  & 95     &  & -1.0 & 11   & 11  & 94     \\
           &           & $\beta_3$ & -1.1 & 27   & 26  & 95     &  & 0.4  & 19   & 18  & 95     \\
           & single001 & $\beta_0$ & 2.4  & 25   & 24  & 95     &  & 0.8  & 17   & 17  & 95     \\
           &           & $\beta_1$ & -0.4 & 27   & 26  & 95     &  & -0.2 & 19   & 18  & 95     \\
           &           & $\beta_2$ & -3.0 & 16   & 15  & 95     &  & -0.9 & 11   & 11  & 94     \\
           &           & $\beta_3$ & -0.2 & 29   & 28  & 95     &  & 0.3  & 20   & 19  & 95     \\
           & ave110    & $\beta_0$ & 2.5  & 25   & 23  & 95     &  & 0.8  & 17   & 17  & 95     \\
           &           & $\beta_1$ & -0.4 & 26   & 25  & 94     &  & -0.1 & 19   & 18  & 94     \\
           &           & $\beta_2$ & -2.9 & 16   & 15  & 95     &  & -1.0 & 11   & 11  & 95     \\
           &           & $\beta_3$ & -0.5 & 29   & 28  & 95     &  & 0.3  & 20   & 19  & 95     \\
           & agg110    & $\beta_0$ & 2.3  & 26   & 24  & 93     &  & 0.5  & 17   & 17  & 95     \\
           &           & $\beta_1$ & -0.4 & 29   & 27  & 93     &  & 0.0  & 20   & 19  & 94     \\
           &           & $\beta_2$ & -2.7 & 17   & 16  & 93     &  & -0.7 & 12   & 11  & 94     \\
           &           & $\beta_3$ & -0.7 & 30   & 28  & 94     &  & 0.2  & 20   & 19  & 95     \\
           & ave111    & $\beta_0$ & 2.5  & 25   & 23  & 95     &  & 0.8  & 17   & 16  & 96     \\
           &           & $\beta_1$ & -0.3 & 27   & 25  & 95     &  & -0.2 & 19   & 18  & 94     \\
           &           & $\beta_2$ & -2.8 & 15   & 15  & 95     &  & -1.0 & 11   & 11  & 95     \\
           &           & $\beta_3$ & -0.6 & 28   & 27  & 95     &  & 0.3  & 19   & 19  & 95     \\
           & agg111    & $\beta_0$ & 2.6  & 26   & 24  & 93     &  & 0.5  & 17   & 17  & 95     \\
           &           & $\beta_1$ & -0.3 & 29   & 27  & 92     &  & 0.0  & 20   & 19  & 94     \\
           &           & $\beta_2$ & -2.6 & 18   & 17  & 94     &  & -0.7 & 13   & 12  & 93     \\
           &           & $\beta_3$ & -1.3 & 28   & 26  & 93     &  & 0.3  & 19   & 18  & 95     \\
           & omn111    & $\beta_0$ & 2.4  & 25   & 23  & 94     &  & 0.6  & 17   & 16  & 95     \\
           &           & $\beta_1$ & -0.4 & 27   & 26  & 93     &  & -0.1 & 19   & 18  & 94     \\
           &           & $\beta_2$ & -2.7 & 16   & 15  & 93     &  & -0.8 & 12   & 11  & 94     \\
           &           & $\beta_3$ & -0.7 & 28   & 26  & 94     &  & 0.3  & 19   & 19  & 95     \\
\bottomrule
\end{tabular}
\end{table}

\begin{table}
\caption{Empirical relative efficiency for eight estimators (defined in Section~\ref{Simulation Studies}) under the combination of different sample sizes ($n$) and correlation values between two longitudinal secondary outcomes ($\rho$). All secondary data are fully observed.}
\label{table2}
\scriptsize
\centering
\begin{tabular}{lccccccccccc}
\toprule
         &         &         &  & \multicolumn{8}{c}{Empirical relative efficiency}                                     \\\cline{5-12} 
         &         &         &  & single100 & single010 & single001 & ave110 & agg110 & ave111 & agg111 & omn111 \\\midrule
n=300 & $\rho$=0   & $\beta_0$ &  & 1.01      & 1.05      & 1.03      & 1.06   & 1.04   & 1.08   & 1.08   & 1.10   \\
    &         & $\beta_1$ &  & 1.10      & 1.01      & 1.08      & 1.18   & 1.21   & 1.15   & 1.21   & 1.21   \\
         &         & $\beta_2$ &  & 1.05      & 1.10      & 1.03      & 1.10   & 1.07   & 1.14   & 1.05   & 1.15   \\
         &         & $\beta_3$ &  & 0.98      & 1.20      & 1.01      & 1.02   & 0.98   & 1.10   & 1.15   & 1.12   \\
         & $\rho$=0.4 & $\beta_0$ &  & 1.03      & 1.07      & 1.03      & 1.07   & 1.01   & 1.11   & 1.08   & 1.11   \\
         &         & $\beta_1$ &  & 1.19      & 1.01      & 1.16      & 1.27   & 1.18   & 1.23   & 1.18   & 1.26   \\
         &         & $\beta_2$ &  & 1.10      & 1.18      & 1.07      & 1.15   & 1.03   & 1.22   & 0.94   & 1.15   \\
         &         & $\beta_3$ &  & 0.98      & 1.27      & 1.00      & 1.01   & 0.97   & 1.14   & 1.20   & 1.16   \\
         & $\rho$=0.8 & $\beta_0$ &  & 1.04      & 1.11      & 1.04      & 1.06   & 0.92   & 1.12   & 1.02   & 1.07   \\
         &         & $\beta_1$ &  & 1.26      & 1.01      & 1.23      & 1.29   & 0.98   & 1.26   & 0.97   & 1.17   \\
         &         & $\beta_2$ &  & 1.13      & 1.20      & 1.11      & 1.15   & 0.87   & 1.22   & 0.71   & 1.01   \\
         &         & $\beta_3$ &  & 0.98      & 1.36      & 0.99      & 1.00   & 0.94   & 1.16   & 1.22   & 1.18   \\\midrule
n=600  & $\rho$=0   & $\beta_0$ &  & 1.04      & 1.04      & 1.06      & 1.08   & 1.10   & 1.09   & 1.13   & 1.14   \\
   &         & $\beta_1$ &  & 1.13      & 1.00      & 1.08      & 1.19   & 1.24   & 1.14   & 1.24   & 1.21   \\
         &         & $\beta_2$ &  & 1.05      & 1.09      & 1.07      & 1.12   & 1.13   & 1.14   & 1.12   & 1.20   \\
         &         & $\beta_3$ &  & 0.99      & 1.19      & 1.03      & 1.03   & 1.02   & 1.12   & 1.18   & 1.16   \\
         & $\rho$=0.4 & $\beta_0$ &  & 1.06      & 1.05      & 1.07      & 1.09   & 1.07   & 1.12   & 1.12   & 1.15   \\
         &         & $\beta_1$ &  & 1.21      & 1.00      & 1.16      & 1.27   & 1.19   & 1.21   & 1.19   & 1.26   \\
         &         & $\beta_2$ &  & 1.12      & 1.14      & 1.12      & 1.18   & 1.11   & 1.21   & 1.02   & 1.21   \\
         &         & $\beta_3$ &  & 0.98      & 1.27      & 1.02      & 1.02   & 1.00   & 1.15   & 1.21   & 1.19   \\
         & $\rho$=0.8 & $\beta_0$ &  & 1.06      & 1.09      & 1.08      & 1.09   & 0.99   & 1.14   & 1.08   & 1.13   \\
         &         & $\beta_1$ &  & 1.26      & 1.00      & 1.22      & 1.28   & 0.99   & 1.25   & 0.98   & 1.17   \\
         &         & $\beta_2$ &  & 1.17      & 1.19      & 1.18      & 1.20   & 0.98   & 1.25   & 0.81   & 1.12   \\
         &         & $\beta_3$ &  & 0.99      & 1.36      & 1.02      & 1.01   & 0.99   & 1.17   & 1.29   & 1.23  \\
\bottomrule
\end{tabular}
\end{table}

\begin{table}
\caption{ Overall evaluations of MinBo for parameter estimation in the main model when working models are misspecified. We consider the combination of different sample sizes ($n$) and correlation values ($\rho$), given three secondary datasets are fully observed. MCSD: Monte Carlo standard deviation; ASE: asymptotic standard error; ERE: empirical relative efficiency; CP: coverage probability. All values (except ERE) are multiplied by $100$.}
\label{table3}
\scriptsize
\centering
\begin{tabular}{cccccccccccccc}
\toprule
        &        &         & \multicolumn{5}{c}{n=300}         &  & \multicolumn{5}{c}{n=600}         \\ \cline{4-8} \cline{10-14} 
        &        &         & Bias & MCSD & ASE & ERE  & 95\%CP &  & Bias & MCSD & ASE & ERE  & 95\%CP \\\midrule
$\rho$=0   & ave111 & $\beta_0$ & 2.5  & 24   & 24  & 1.06 & 94     &  & 1.2  & 17   & 17  & 1.07 & 95     \\
        &        & $\beta_1$ & -0.8 & 27   & 26  & 1.13 & 94     &  & -0.6 & 19   & 18  & 1.12 & 95     \\
        &        & $\beta_2$ & -2.7 & 16   & 15  & 1.08 & 94     &  & -1.1 & 11   & 11  & 1.08 & 94     \\
        &        & $\beta_3$ & -0.4 & 28   & 27  & 1.06 & 95     &  & 0.1  & 19   & 19  & 1.07 & 95     \\
        & ave101 & $\beta_0$ & 2.6  & 25   & 24  & 1.05 & 94     &  & 1.2  & 17   & 17  & 1.06 & 95     \\
        &        & $\beta_1$ & -0.9 & 27   & 26  & 1.15 & 94     &  & -0.6 & 18   & 18  & 1.15 & 95     \\
        &        & $\beta_2$ & -2.7 & 16   & 15  & 1.09 & 94     &  & -1.1 & 11   & 11  & 1.10 & 94     \\
        &        & $\beta_3$ & -0.4 & 29   & 28  & 0.98 & 95     &  & 0.1  & 20   & 20  & 1.00 & 95     \\
        & agg111 & $\beta_0$ & 2.4  & 24   & 23  & 1.08 & 94     &  & 1.1  & 16   & 16  & 1.12 & 95     \\
        &        & $\beta_1$ & -0.8 & 26   & 25  & 1.22 & 94     &  & -0.5 & 18   & 17  & 1.23 & 95     \\
        &        & $\beta_2$ & -2.7 & 16   & 15  & 1.08 & 93     &  & -1.1 & 11   & 10  & 1.14 & 94     \\
        &        & $\beta_3$ & -0.3 & 27   & 26  & 1.11 & 94     &  & 0.0  & 19   & 18  & 1.14 & 94     \\
        & agg101 & $\beta_0$ & 2.6  & 25   & 23  & 1.03 & 93     &  & 1.2  & 17   & 16  & 1.08 & 96     \\
        &        & $\beta_1$ & -0.9 & 26   & 25  & 1.21 & 94     &  & -0.5 & 18   & 17  & 1.23 & 95     \\
        &        & $\beta_2$ & -2.8 & 16   & 15  & 1.08 & 93     &  & -1.2 & 11   & 10  & 1.14 & 94     \\
        &        & $\beta_3$ & -0.3 & 29   & 28  & 0.94 & 94     &  & 0.1  & 20   & 20  & 0.98 & 95     \\
        & omn111 & $\beta_0$ & 2.5  & 24   & 23  & 1.08 & 94     &  & 1.2  & 16   & 16  & 1.11 & 96     \\
        &        & $\beta_1$ & -0.9 & 26   & 25  & 1.21 & 94     &  & -0.6 & 18   & 18  & 1.20 & 95     \\
        &        & $\beta_2$ & -2.7 & 16   & 15  & 1.09 & 94     &  & -1.2 & 11   & 11  & 1.13 & 94     \\
        &        & $\beta_3$ & -0.3 & 28   & 26  & 1.06 & 95     &  & 0.1  & 19   & 19  & 1.09 & 95     \\\midrule
$\rho$=0.4 & ave111 & $\beta_0$ & 2.3  & 24   & 23  & 1.09 & 95     &  & 0.9  & 17   & 16  & 1.09 & 95     \\
        &        & $\beta_1$ & -0.4 & 26   & 25  & 1.19 & 94     &  & -0.1 & 18   & 18  & 1.18 & 95     \\
        &        & $\beta_2$ & -2.6 & 15   & 15  & 1.12 & 95     &  & -1.1 & 11   & 11  & 1.13 & 94     \\
        &        & $\beta_3$ & -0.6 & 27   & 27  & 1.09 & 95     &  & 0.2  & 19   & 19  & 1.10 & 96     \\
        & ave101 & $\beta_0$ & 2.5  & 24   & 23  & 1.06 & 95     &  & 0.9  & 17   & 16  & 1.08 & 95     \\
        &        & $\beta_1$ & -0.5 & 26   & 25  & 1.24 & 94     &  & -0.1 & 18   & 17  & 1.23 & 94     \\
        &        & $\beta_2$ & -2.6 & 15   & 15  & 1.14 & 95     &  & -1.1 & 11   & 11  & 1.16 & 94     \\
        &        & $\beta_3$ & -0.7 & 29   & 28  & 0.99 & 95     &  & 0.1  & 20   & 20  & 1.00 & 95     \\
        & agg111 & $\beta_0$ & 2.2  & 24   & 22  & 1.10 & 95     &  & 0.6  & 16   & 16  & 1.14 & 94     \\
        &        & $\beta_1$ & -0.4 & 25   & 24  & 1.26 & 93     &  & 0.1  & 18   & 17  & 1.27 & 95     \\
        &        & $\beta_2$ & -2.5 & 16   & 15  & 1.09 & 94     &  & -1.1 & 11   & 10  & 1.16 & 93     \\
        &        & $\beta_3$ & -0.5 & 26   & 25  & 1.18 & 95     &  & 0.1  & 18   & 18  & 1.21 & 95     \\
        & agg101 & $\beta_0$ & 2.6  & 24   & 23  & 1.02 & 94     &  & 0.8  & 17   & 16  & 1.08 & 95     \\
        &        & $\beta_1$ & -0.7 & 25   & 24  & 1.26 & 93     &  & 0.0  & 18   & 17  & 1.27 & 95     \\
        &        & $\beta_2$ & -2.6 & 16   & 15  & 1.09 & 94     &  & -1.1 & 11   & 10  & 1.17 & 93     \\
        &        & $\beta_3$ & -0.7 & 29   & 28  & 0.95 & 95     &  & 0.1  & 20   & 20  & 0.98 & 95     \\
        & omn111 & $\beta_0$ & 2.4  & 24   & 23  & 1.10 & 94     &  & 0.8  & 16   & 16  & 1.14 & 95     \\
        &        & $\beta_1$ & -0.6 & 25   & 24  & 1.29 & 94     &  & -0.1 & 18   & 17  & 1.28 & 94     \\
        &        & $\beta_2$ & -2.6 & 15   & 15  & 1.12 & 94     &  & -1.2 & 11   & 10  & 1.17 & 94     \\
        &        & $\beta_3$ & -0.6 & 27   & 26  & 1.10 & 95     &  & 0.1  & 19   & 18  & 1.13 & 95     \\\midrule
$\rho$=0.8 & ave111 & $\beta_0$ & 2.3  & 24   & 23  & 1.10 & 95     &  & 0.7  & 17   & 16  & 1.11 & 95     \\
        &        & $\beta_1$ & -0.3 & 26   & 25  & 1.23 & 95     &  & 0.0  & 18   & 17  & 1.22 & 94     \\
        &        & $\beta_2$ & -2.5 & 15   & 15  & 1.14 & 95     &  & -1.0 & 11   & 10  & 1.16 & 95     \\
        &        & $\beta_3$ & -0.6 & 27   & 26  & 1.11 & 95     &  & 0.3  & 19   & 19  & 1.13 & 95     \\
        & ave101 & $\beta_0$ & 2.4  & 24   & 23  & 1.05 & 94     &  & 0.7  & 17   & 16  & 1.08 & 95     \\
        &        & $\beta_1$ & -0.4 & 25   & 24  & 1.27 & 94     &  & 0.0  & 18   & 17  & 1.26 & 94     \\
        &        & $\beta_2$ & -2.4 & 15   & 15  & 1.15 & 95     &  & -1.0 & 11   & 10  & 1.19 & 94     \\
        &        & $\beta_3$ & -0.5 & 29   & 28  & 0.99 & 95     &  & 0.2  & 20   & 20  & 1.00 & 95     \\
        & agg111 & $\beta_0$ & 2.3  & 24   & 22  & 1.08 & 93     &  & 0.4  & 16   & 16  & 1.13 & 94     \\
        &        & $\beta_1$ & -0.2 & 26   & 25  & 1.17 & 93     &  & 0.4  & 19   & 17  & 1.16 & 93     \\
        &        & $\beta_2$ & -2.2 & 16   & 15  & 1.01 & 93     &  & -0.8 & 11   & 10  & 1.11 & 94     \\
        &        & $\beta_3$ & -1.1 & 26   & 24  & 1.24 & 94     &  & 0.1  & 18   & 17  & 1.29 & 94     \\
        & agg101 & $\beta_0$ & 2.4  & 25   & 23  & 0.98 & 93     &  & 0.5  & 17   & 16  & 1.04 & 95     \\
        &        & $\beta_1$ & -0.5 & 26   & 25  & 1.17 & 93     &  & 0.2  & 19   & 17  & 1.17 & 94     \\
        &        & $\beta_2$ & -2.2 & 16   & 15  & 1.02 & 93     &  & -0.9 & 11   & 10  & 1.12 & 94     \\
        &        & $\beta_3$ & -0.7 & 30   & 28  & 0.94 & 94     &  & 0.1  & 20   & 20  & 0.98 & 95     \\
        & omn111 & $\beta_0$ & 2.3  & 24   & 22  & 1.09 & 94     &  & 0.6  & 16   & 16  & 1.14 & 95     \\
        &        & $\beta_1$ & -0.4 & 25   & 24  & 1.27 & 93     &  & 0.1  & 18   & 17  & 1.25 & 94     \\
        &        & $\beta_2$ & -2.4 & 15   & 15  & 1.09 & 93     &  & -0.9 & 11   & 10  & 1.17 & 94     \\
        &        & $\beta_3$ & -0.8 & 27   & 26  & 1.13 & 94     &  & 0.1  & 18   & 18  & 1.17 & 94    \\\bottomrule
\end{tabular}
\end{table}

\begin{table}
\caption{Estimated relative efficiency between MLE without considering secondary data and estimators based on different data integration schemes}
\label{table4}
\centering
\footnotesize
\begin{tabular}{lccccccc}
\toprule
          & single100 & single010 & single001 & ave111 & agg111 & omn111 \\\midrule
Intercept & 1.15      & 1.06      & 1.68      & 1.39   & 1.81   & 1.69   \\
BMI       & 1.06      & 1.05      & 1.07      & 1.09   & 1.14   & 1.12   \\
Drink     & 1.14      & 2.62      & 1.00      & 1.96   & 2.81   & 2.87   \\
Smoking   & 1.10      & 2.87      & 1.01      & 2.05   & 2.88   & 3.03   \\
Hemoglobin       & 1.16      & 1.08      & 2.03      & 1.53   & 2.15   & 2.00   \\
Age       & 1.13      & 1.02      & 1.04      & 1.08   & 1.18   & 1.12   \\\bottomrule
\end{tabular}
\end{table}

\begin{table}
\caption{Summary for estimators agg111 and omn111. ASE: asymptotic standard errors; ERE: estimated relative efficiency compared to MLE; OR: odds ratio; LL: lower limit; UL: upper limit; P-value\_A: P-values for agg111; P-value\_O: P-values for omn111; P-value\_MLE: P-values for MLE.}
\label{table5}
\footnotesize
\centering
\begin{tabular}{lcccccccc}
        \toprule
        & \multicolumn{8}{c}{agg111}  \\ \cline{2-9} 
  & Estimates & ASE   & ERE   & OR    & LL    & UL    & P-value\_A & P-value\_MLE \\\midrule
Intercept  & -5.165    & 0.880 & 1.805 & 0.006 & 0.001 & 0.032 & 0.000      & 0.000        \\
BMI        & 0.046     & 0.015 & 1.142 & 1.047 & 1.016 & 1.078 & 0.003      & 0.011        \\
Drink      & 0.214     & 0.077 & 2.808 & 1.239 & 1.066 & 1.441 & 0.005      & 0.368        \\
Smoking    & -0.126    & 0.086 & 2.880 & 0.882 & 0.745 & 1.044 & 0.143      & 0.668        \\
Hemoglobin & 0.027     & 0.041 & 2.146 & 1.027 & 0.948 & 1.113 & 0.510      & 0.488        \\
Age        & 0.055     & 0.010 & 1.181 & 1.056 & 1.035 & 1.078 & 0.000      & 0.000        \\\toprule
           & \multicolumn{8}{c}{omn111}                                    \\ \cline{2-9} 
       & Estimates & ASE   & ERE   & OR    & LL    & UL    & P-value\_O & P-value\_MLE \\
       \midrule
Intercept  & -4.831    & 0.910 & 1.687 & 0.008 & 0.001 & 0.047 & 0.000      & 0.000        \\
BMI        & 0.043     & 0.015 & 1.118 & 1.043 & 1.012 & 1.075 & 0.006      & 0.011        \\
Drink      & 0.174     & 0.076 & 2.869 & 1.190 & 1.025 & 1.381 & 0.022      & 0.368        \\
Smoking    & -0.126    & 0.084 & 3.030 & 0.882 & 0.748 & 1.039 & 0.133      & 0.668        \\
Hemoglobin & 0.033     & 0.042 & 2.002 & 1.034 & 0.951 & 1.124 & 0.432      & 0.488        \\
Age        & 0.048     & 0.011 & 1.121 & 1.049 & 1.028 & 1.071 & 0.000      & 0.000       
\\\bottomrule    
\end{tabular}
\end{table}

\end{document}

\label{firstpage}

 \begin{center}
    {\LARGE {Supporting Information for "An Efficient Data Integration Scheme for Synthesizing Information from Multiple Secondary {Datasets} for the {Parameter Inference of the Main Analysis}" by Chixiang Chen, Ming Wang, Shuo Chen}}
\end{center}

%  \begin{center}
%     {\LARGE
%     Chixiang Chen$^{1*}$, Ming Wang$^{2}$, Shuo Chen$^{1}$\\}
% \end{center}
%   \begin{center}
%     $^{1}$ Division of Biostatistics and Bioinformatics,\\ University of Maryland School of Medicine, MD, USA.\\ 
%     $^{2}$ \revision{Department of Population and Quantitative Health Sciences,School of Medicine, Case Western Reserve University, Cleveland, OH, USA}\\ 
%     $^{*}$Contact: chixiang.chen@som.umaryland.edu\\
% \end{center}

%\appendix

In this supplementary material, we will present regularity conditions (Section \ref{Regularity conditions}), the proofs for theorems and properties from the main text of paper (Section \ref{Proof  for Theorem 1}, \ref{Proof  for Theorem 2}, and \ref{Other forms}), extra simulation results (Section \ref{Extra simulation results}), the extension to the missing-data problem (Section \ref{Extension to missing-data problem}), \revision{examples of over-identified estimating functions (Section \ref{Over-identified estimating functions}), \revisiontwo{and more discussion on estimation efficiency} (Section \ref{Discussion based on Semi-parametric theory}).}
The proofs are built based upon classic moment conditions, which are widely adopted in method of moment \cite{newey1994large} and empirical likelihood \cite{qin1994empirical}. The notation $\lVert\cdot\lVert$ represents $L2$ norm.

\section{Regularity conditions}\label{Regularity conditions}
\begin{condition}\label{c1}
	 $E\big\{\bfg(\bfD_{0i};\bfbeta)\big\}=\bfzero$ if and only if $\bfbeta=\bfbeta_0$.
\end{condition}

\begin{condition}\label{c2}
	For $k=1,\ldots,K$, there exist values of parameters $\bftheta_k^\ast$ such that $E\big\{\bfh_k(\bfD_{ki};\bftheta_k^\ast)\big\}=\bfzero$.
\end{condition}

\begin{condition}\label{c4}
	The functions $\bfg(\bfD_{0i};\bfbeta)$ is twice continuously differentiable, $E\lVert \bfg(\bfD_{0i};\bfbeta_0)\lVert^2$ is finite, and $\lVert\partial^2\bfg(\bfD_{0i};\bfbeta)/\partial\bfbeta\partial\bfbeta^T\lVert$ can be bounded by some integrable function in the neighborhood of $\bfbeta_0$.
\end{condition}

\begin{condition}\label{c5}
    For $k=1,\ldots,K$ and $\bftheta_k^\ast$ defined in Condition \ref{c1}, {$E\big\{\bfh_k(\bfD_{ki}; \bftheta^\ast_k)\bfh_k^T(\bfD_{ki}; \bftheta^\ast_{k})\big\}$} is positive definite, and $\{\partial^2 \bfh_k(\bfD_{ki}; \bftheta_{k})\}/(\partial\bftheta_k^T\partial \bftheta_k)$ is continuous in the neighborhood of $\bftheta^\ast_k$. Furthermore, $\lVert \{\partial\bfh_k(\bfD_{ki}; \bftheta_k)\}/(\partial\bftheta_k^T)\rVert$, $\lVert \{\partial^2 \bfh_k(\bfD_{ki}; \bftheta_k)\}/(\partial\bftheta_k^T \partial\bftheta_k)\rVert$, and $\lVert \bfh_k(\bfD_{ki}; \bftheta_k)\rVert^3$ are bounded by some integrable function around $\bftheta^\ast_k$. 
\end{condition}

 Condition \ref{c1} and \ref{c2} facilitate the estimation consistency for the main parameter of interest. Condition \ref{c4} is a regularity to guarantee a valid Taylor expansion \citep{newey1994large}. Condition $\ref{c5}$ contains extra moment conditions adopted for valid empirical likelihood inference \citep{qin1994empirical}. The following two sections will provide detailed proofs for Theorem 1 and Theorem 2 of the main paper. The proof for Theorem 3 can be similarly derived by combining proofs of Theorem 1 and Theorem 2. Thus, we will omit the details.

\section{Proof for Theorem 1}\label{Proof  for Theorem 1}
In this proof, we consider a general case where there are K \revision{secondary} data sets. In order to solve the problem of constrained maximization in (3) \revision{of} the main paper, we employ the empirical likelihood theory to \revision{facilitate} the proof \citep{qin1994empirical}. To be specific, first, Lagrange's multipliers $\bflambda_k$ \revisiontwo{are} introduced into the optimization procedure. Thereafter, by maximizing $\prod_{i=1}^np_{ki}$ \revision{with respect to $p_{ki}$ and $\bftheta_k$}, the estimated \revision{weights} can be solved by $\hat{p}_{ki}=(1/n)/\{1+\hat{\bflambda}^T_kR_{ki}\bfh_k(\bfD_{ki}; \hat{\bftheta}_k)\}$, where the estimated Lagrange's multipliers $\hat{\bflambda}_k$ and empirical likelihood estimator $\hat{\bftheta}_k$ \revision{can} be obtained by solving the following estimating equations
\begin{equation}\label{lambda and theta}
\begin{split}
        \frac{1}{n}&\sum_{i=1}^n\frac{R_{ki}\bfh_k(\bfD_{ki};{\bftheta}_k)}{1+\bflambda_kR_{ki}\bfh_k(\bfD_{ki};{\bftheta}_k)}=\bfzero;\\
        \frac{1}{n}&\sum_{i=1}^n\frac{R_{ki}\bfh_k(\bfD_{ki};{\bftheta}_k)}{1+\bflambda_kR_{ki}\bfh_k(\bfD_{ki};{\bftheta}_k)}\bigg\{\frac{\partial R_{ki}\bfh_k(\bfD_{ki};\bftheta_k)}{\partial\bftheta^T}\bigg\}^T\bflambda_k=\bfzero.
\end{split}
\end{equation}
By expanding the equations in (\ref{lambda and theta}) at $\bflambda_k=\bfzero$ and $\bftheta_k^\ast$ under Condition \ref{c5}, we derive the asymptotic representation of $\hat{\bflambda}_k$, i.e., for each $k=1,\ldots,K$,
\begin{equation}\label{ap1}
\hat{\bflambda}_k=\frac{1}{n}\bfS_k\bfQ_k(\bftheta^\ast_k)+o_p(\bfn^{-\frac{1}{2}}),
\end{equation}
with $\bfQ_k(\bftheta^\ast_k)=\sum_{i=1}^{n} R_{ki}\bfh_k(\bfD_{ki}; \bftheta^\ast_k)$. The notation $\bfS_k$ is defined in the Theorem 1 \revision{of} the mainn paper. On the other hand, by Taylor expansion again under Condition \ref{c5}, the estimated \revision{weights} for each $k$ \revision{are} further expressed by $\hat{p}_{ki}=(1/n)[1-\hat{\bflambda}_k^T R_{ki}\bfh_k(\bfD_{ki}; \bftheta^\ast_k) \{1+o_p(1)\}]$. \revision{Together with the estimating equations in (6) of the main paper}, we have
\begin{equation*}
\begin{split}
\bfzero=&\sum_{i=1}^{n}\bar{p}_{i} \bfg(\bfD_{0i};\bar{\bfbeta})\\
=&\frac{1}{n}\sum_{i=1}^{n} \bfg(\bfD_{0i};\bar{\bfbeta}) \left[1-\sum_{k=1}^K\omega_k\hat{\bflambda}^T_k R_{ki}\bfh_k(\bfD_{ki}; \bftheta^\ast_k)\big\{1+o_p(1)\big\} \right]\\
=&\frac{1}{n}\sum_{i=1}^{n} \bfg(\bfD_{0i};{\bfbeta_0})+ \frac{1}{n}\sum_{i=1}^{n} \frac{\partial \bfg(\bfD_{0i};{\bfbeta}_{0})}{\partial\bfbeta^T} (\bar{\bfbeta}-\bfbeta_0)\\
&-\frac{1}{n}\sum_{i=1}^{n}\sum_{k=1}^K\omega_k R_{ki}\bfg(\bfD_{0i};{\bfbeta_0})\bfh_k(\bfD_{ki}; \bftheta^\ast_k)^T\hat{\bflambda}_k+o_p(\bfn^{-\frac{1}{2}}).
\end{split}
\end{equation*}
The second equality holds by replacing $\bar{p}_i$ with $\sum_{i=1}^K\omega_k\hat{p}_{ki}$ and the fact that $\sum_{i=1}^K\omega_k=1$. The third equation is based on Taylor expansion with respect to $\bfbeta_0$ under Condition \ref{c4}. Thus, the asymptotic expansion of the estimator $\bar{\bfbeta}$ can be written as
\begin{equation*}
\begin{split}
&n^{\frac{1}{2}}(\bar{\bfbeta}-\bfbeta_0)\\
=&-\left(\frac{1}{n} \sum_{i=1}^n \frac{\partial \bfg(\bfD_{0i};{\bfbeta}_{0})}{\partial\bfbeta^T}\right)^{-1} \bigg[  n^{-\frac{1}{2}}\sum_{i=1}^n \bfg(\bfD_{0i};{\bfbeta}_0)\\
&- n^{-\frac{1}{2}} \sum_{i=1}^{n}\sum_{k=1}^K \omega_k\Big\{ R_{ki}\bfg(\bfD_{0i};{\bfbeta}_0) \bfh_k(\bfD_{ki}; \bftheta^\ast_k)^T \Big\} \hat{\bflambda}_k \bigg]
+o_p(\bfone)\\
=& -\bfvarGamma^{-1}\left\{n^{-\frac{1}{2}}\sum_{i=1}^n \bfg(\bfD_{0i};{\bfbeta}_0)  - n^{-\frac{1}{2}}\sum_{k=1}^K\omega_k\bfvarLambda_k\bfS_k\bfQ_{k}(\bftheta^\ast_k)\right\}+o_p(\bfone).
\end{split}
\end{equation*}

Finally, by simple matrix algebra, we can obtain {the asymptotic normality of $n^{1/2}(\bar{\bfbeta}-\bfbeta_0)$,} where the asymptotic covariance matrix in Theorem 1 \revision{of} the main paper could be easily \revision{calculated} based on the influence function derived above \revision{and by setting $K=2$}.\newline

\section{Proof  for Theorem 2}\label{Proof  for Theorem 2}
In this proof, we consider a general case where there are K auxiliary data sets. \revision{Based on (8) \revision{of} the main paper, the asymptotic expansion of estimated \revision{weights} $\hat{p}_{ki}$, and Lagrange multipliers $\hat{\bflambda}_k$ derived in Section \ref{Proof  for Theorem 1},}  we have
\begin{equation*}
\begin{split}
\bfzero=&\sum_{i=1}^{n}\tilde{p}_{i} \bfg(\bfD_{0i}^u;\tilde{\bfbeta})\\
=&\frac{1}{n}\sum_{i=1}^{n} \bfg(\bfD_{0i};\tilde{\bfbeta}) \left[1-\sum_{k=1}^K\hat{\bflambda}^T_k R_{ki}\bfh_k(\bfD_{ki}; \bftheta^\ast_k)\big\{1+o_p(1)\big\} \right]+o_p\bigg(\frac{1}{n}\sum_{i=1}^{n} \bfg(\bfD_{0i};\tilde{\bfbeta})\bigg)\\
=&\frac{1}{n}\sum_{i=1}^{n} \bfg(\bfD_{0i};{\bfbeta_0})+ \frac{1}{n}\sum_{i=1}^{n} \frac{\partial \bfg(\bfD_{0i};{\bfbeta}_{0})}{\partial\bfbeta^T} (\tilde{\bfbeta}-\bfbeta_0)\\
&-\frac{1}{n}\sum_{i=1}^{n}\sum_{k=1}^K R_{ki}\bfg(\bfD_{0i};{\bfbeta_0})\bfh_k(\bfD_{ki}; \bftheta^\ast_k)^T\hat{\bflambda}_k+o_p(\bfn^{-\frac{1}{2}}),
\end{split}
\end{equation*}
The second equality holds by replacing $\tilde{p}_i$ with $\prod_{i=1}^K\hat{p}_{ki}$ and the fact that 
\begin{equation*}
        \max_{i=1,\ldots,n}\hat{\bflambda}^T_k R_{ki}\bfh_k(\bfD_{ki}; \bftheta^\ast_k)=o_p(1),
\end{equation*}
\revision{which is based on the empirical likelihood theory} under Condition \ref{c5} \citep{owen2001empirical}. The third equation is based on Taylor expansion with respect to $\bfbeta_0$ under Condition \ref{c4}. Thus, the asymptotic expansion of the estimator $\tilde{\bfbeta}$ \revision{can be} written as
\begin{equation*}
\begin{split}
&n^{\frac{1}{2}}(\tilde{\bfbeta}-\bfbeta_0)\\
=&-\left(\frac{1}{n} \sum_{i=1}^n \frac{\partial \bfg(\bfD_{0i};{\bfbeta}_{0})}{\partial\bfbeta^T}\right)^{-1} \bigg[  n^{-\frac{1}{2}}\sum_{i=1}^n \bfg(\bfD_{0i};{\bfbeta}_0)\\
&- n^{-\frac{1}{2}} \sum_{i=1}^{n}\sum_{k=1}^K \Big\{ R_{ki}\bfg(\bfD_{0i};{\bfbeta}_0) \bfh_k(\bfD_{ki}; \bftheta^\ast_k)^T \Big\} \hat{\bflambda}_k \bigg]
+o_p(\bfone)\\
=& -\bfvarGamma^{-1}\left\{n^{-\frac{1}{2}}\sum_{i=1}^n \bfg(\bfD_{0i};{\bfbeta}_0)  - n^{-\frac{1}{2}}\sum_{k=1}^K\bfvarLambda_k\bfS_k\bfQ_{k}(\bftheta^\ast_k)\right\}+o_p(\bfone).
\end{split}
\end{equation*}

Finally, by simple matrix algebra, we can obtain {the asymptotic normality of $n^{1/2}(\tilde{\bfbeta}-\bfbeta_0)$,} \revision{where the asymptotic covariance matrix in Theorem 2 \revision{of} the main paper can be easily calculated based on the influence function derived above and by setting $K=2$.}\newline

The proof for Theorem 3 in the main paper can be similarly derived by combining the proofs from Theorem 1 and Theorem 2. We omit the details here.

\section{Other estimation forms}\label{Other forms}
In Section 2 \revision{of the main paper}, we propose a re-weighted estimation (2) and (3), which is the same to the re-weighted estimation introduced in \cite{chen2021improving}. \revision{Briefly, \cite{chen2021improving} assumes that the} first $m_1$ subjects have secondary data, and rest of them have no secondary record. Then, the parameter $\bfbeta$ can be solved by 
\begin{equation}\label{equivalent ee}
    \sum_{i=1}^{m_1}m_i\hat{p}^\ast_{ki}\bfg(\bfD_{0i};\bfbeta)+\sum_{i=m_1+1}^{n}\bfg(\bfD_{0i};\bfbeta)=\bfzero,
\end{equation}
where the non-negative weights $\hat{p}^\ast_{ki}$ are solved by maximizing $\prod_{i=1}^{m_1}p_{ki}$ under the following \revision{constraints}
\begin{equation}\label{equivalent constrain}
    \sum_{i=1}^{m_1}{p}^\ast_{ki}=1,\sum_{i=1}^{m_i}{p}^\ast_{ki}\bfh(\bfD_{ki};\bftheta_k)=\bfzero.
\end{equation}

To see \revision{the equivalence between two estimation forms, we first }notice that the estimated weights in (2) of the main paper are equal to $\hat{p}_{ki}=n^{-1}/\{1+\hat{\bflambda}_k^TR_{ki}\bfh_k(\bfD_{ki};\hat{\bftheta}_k)\}$ by applying the Lagrange multiplier technique \citep{qin1994empirical} for $i=1,\ldots,m_1$. These weights will become $1/n$ when $i=m_1+1,\ldots,n$. The \revision{constraints} in (3) from the main paper then become $\sum_{i=1}^{m_1}p_{ki}=m_1/n$ and $\sum_{i=1}^{m_1}p_{ki}\bfh_k(\bfD_{ki};\bftheta_k)=\bfzero$, and the estimating equation in (2) from the main paper becomes $\sum_{i=1}^{m_1}\hat{p}_{ki}\bfg(\bfD_{0i};\bfbeta)+\sum_{i=m_1+1}^n(1/n)\bfg(\bfD_{0i};\bfbeta)=\bfzero$. Thus, (2) and (3) from the main paper are reduced to (\ref{equivalent ee}) and (\ref{equivalent constrain}) above by noting the fact that $\hat{p}_{ki}=(m_1/n)\hat{p}^\ast_{ki}$.

\revisiontwo{In addition to (\ref{equivalent ee}) and (\ref{equivalent constrain}), the proposed estimation procedure (2) and (3) of the main paper is asymptotically equivalent to the following empirical likelihood estimation, i.e., to maximize $\prod_{i=1}^np_{ki}$ with respect to $p_{ki}$, $\bfbeta$, $\bftheta_k$, and subject to three {constraints}
\begin{equation}
    \sum_{i=1}^np_{ki}=1, \sum_{i=1}^np_{ki}\bfg(\bfD_{0i};\bfbeta)=\bfzero,
    \sum_{i=1}^np_{ki}R_{ki}\bfh_k(\bfD_{ki};\bftheta_k)=\bfzero.
\end{equation}

To show asymptotic equivalence, we apply the results from \cite{qin1994empirical} and introduce two Lagrange multipliers $\bflambda_{k1}$ and $\bflambda_{k2}$ corresponding to above two moment {constraints}. In the following derivation, we suppress the subscript $k$ for ease of notation. It is easy to check that $\hat{\bflambda}_{1}$ and $\hat{\bflambda}_{2}$ are the solutions to the following equations \citep{qin1994empirical}
\begin{equation}\label{joint profiling lambdas}
    \begin{split}
        \bfQ_{1n}=\frac{1}{n}\sum_{i=1}^n\frac{\bfg(\bfD_{0i};\bfbeta)}{1+\bflambda_1^T\bfg(\bfD_{0i};\bfbeta)+\bflambda_2^TR_i\bfh(\bfD_i;\bftheta)}=&\bfzero\\
        \bfQ_{2n}=\frac{1}{n}\sum_{i=1}^n\frac{R_i\bfh(\bfD_i;\bftheta)}{1+\bflambda_1^T\bfg(\bfD_{0i};\bfbeta)+\bflambda_2^TR_i\bfh(\bfD_i;\bftheta)}=&\bfzero,
    \end{split}
\end{equation}
and $\hat{\bfbeta}$ and $\hat{\bftheta}$ are the solutions to 
\begin{equation} \label{joint profiling beta and theta}
    \begin{split}
        \bfG_{1n}=\frac{1}{n}\sum_{i=1}^n\frac{\big\{\partial \bfg(\bfD_{0i};\bfbeta)/\partial\bfbeta^T\big\}^T\bflambda_1}{1+\bflambda_1^T\bfg(\bfD_{0i};\bfbeta)+\bflambda_2^TR_i\bfh(\bfD_i;\bftheta)}=&\bfzero\\
        \bfG_{2n}=\frac{1}{n}\sum_{i=1}^n\frac{\big\{\partial R_i\bfh(\bfD_i;\bftheta)/\partial\bftheta^T\big\}^T\bflambda_2}{1+\bflambda_1^T\bfg(\bfD_{0i};\bfbeta)+\bflambda_2^TR_i\bfh(\bfD_i;\bftheta)}=&\bfzero.
    \end{split}
\end{equation}

By applying Taylor expansion to (\ref{joint profiling lambdas}) and (\ref{joint profiling beta and theta}) at $\bfbeta_0$, $\bftheta^\ast$, $\bflambda_1=\bfzero$, and $\bflambda_2=\bfzero$, we have
\begin{equation*}
    \begin{split}
        \bfzero=&\bfQ_{1n}+\frac{\partial \bfQ_{1n}}{\partial\bfbeta^T}(\hat{\bfbeta}-\bfbeta_0)+\frac{\partial \bfQ_{1n}}{\partial\bflambda_1^T}\hat{\bflambda}_1+\frac{\partial \bfQ_{1n}}{\partial\bflambda_2^T}\hat{\bflambda}_2+o_p(\bfn^{-1/2})\\
        \bfzero=&\bfQ_{2n}+\frac{\partial \bfQ_{2n}}{\partial\bftheta^T}(\hat{\bftheta}-\bftheta^\ast)+\frac{\partial \bfQ_{2n}}{\partial\bflambda_1^T}\hat{\bflambda}_1+\frac{\partial \bfQ_{2n}}{\partial\bflambda_2^T}\hat{\bflambda}_2+o_p(\bfn^{-1/2})\\
        \bfzero=&\bfG_{1n}+\frac{\partial \bfG_{1n}}{\partial\bflambda_1^T}\hat{\bflambda}_1+\frac{\partial \bfG_{1n}}{\partial\bflambda_2^T}\hat{\bflambda}_2+o_p(\bfn^{-1/2})\\
        \bfzero=&\bfG_{2n}+\frac{\partial \bfG_{2n}}{\partial\bflambda_1^T}\hat{\bflambda}_1+\frac{\partial \bfG_{2n}}{\partial\bflambda_2^T}\hat{\bflambda}_2+o_p(\bfn^{-1/2}),
    \end{split}
\end{equation*}
which hold by the fact that $\partial \bfQ_{1n}/\partial\bftheta^T$, $\partial \bfQ_{2n}/\partial\bfbeta^T$, $\partial \bfG_{1n}/\partial\bfbeta^T$, $\partial \bfG_{1n}/\partial\bftheta^T$, $\partial \bfG_{2n}/\partial\bfbeta^T$, and $\partial \bfG_{2n}/\partial\bftheta^T$ are all zero matrices evaluated at $\bfbeta_0$, $\bftheta^\ast$, $\bflambda_1=\bfzero$, and $\bflambda_2=\bfzero$. By solving the above system, we can obtain the asymptotic expansion of main parameter estimates
\begin{equation}\label{asymptotic expansion}
    n^{1/2}(\hat{\bfbeta}-\bfbeta_0)=-\frac{\partial \bfQ_{1n}}{\partial\bfbeta^T}(n^{1/2}\bfQ_{1n}+n^{1/2}\bfB\bfD^{-1}\bfZ)+o_p(\bfone),
\end{equation}
where $\bfB=(\partial \bfQ_{1n}/\partial\bflambda_2^T, \partial \bfQ_{1n}/\partial\bflambda_1^T,\bfzero)$, $\bfZ=(-\bfQ_{2n}^T,-\bfG_{1n}^T,-\bfG_{2n}^T)^T$, and 
$$
\bfD=\begin{bmatrix} 
	\frac{\partial \bfQ_{2n}}{\partial\bflambda_2^T} & \frac{\partial \bfQ_{2n}}{\partial\bflambda_1^T} & \frac{\partial \bfQ_{2n}}{\partial\bftheta^T} \\
	\bfzero & \frac{\partial \bfG_{1n}}{\partial\bflambda_1^T} & \bfzero\\
	\frac{\partial \bfG_{2n}}{\partial\bflambda_2^T} & \bfzero & \bfzero.
\end{bmatrix}
$$
After some algebra, the term $\bfB\bfD^{-1}\bfZ$ eventually becomes $-(1/n)\bfLambda \bfS \bfQ_{m_1}(\bftheta^\ast)$ defined in (2) from the Supplementary Material \citep{chen2021improving}, given that the first $m_1$ subjects in the study have secondary data, and the rest subjects have no secondary records. Thus, the expression in (\ref{asymptotic expansion}) is also equal to the asymptotic expansion of the estimator $\hat{\bfbeta}_{EN}$ in this article. Moreover, the above empirical likelihood estimator is semi-parametric efficient given the information of $E\{\bfg(\bfD_{0i};\bfbeta)\}=\bfzero$ and $E\{R_{ki}\bfh_k(\bfD_{ki};\bftheta_k)\}=\bfzero$ by following the proof strategy in Section 6.4 of \cite{qin2017biased}. Thus, our proposed estimation scheme (2) and (3) from the main paper is also semi-parametric efficient. Compared to (\ref{joint profiling lambdas}) and (\ref{joint profiling beta and theta}), our proposed estimation scheme requires lower dimension of functions and tuning parameters in the empirical likelihood framework and thus would be numerically more stable and require less computation. In addition, the semiparametric efficient estimator can be conceptually constructed for the scenario with multiple secondary outcomes ($k=1,\ldots, K$) available via our weighted estimation or empirical likelihood by replacing the over-identified function $R_{ki}\bfh_k(\bfD_{ki};\bftheta_k)$ with an enlarged over-identified function
$$\bfh(\bfD_{1i},\ldots,\bfD_{Ki},R_{1i}\ldots,R_{Ki};\bftheta_1,\ldots,\bftheta_K)=(R_{1i}\bfh_1^T(\bfD_{1i};\bftheta_1),\ldots,R_{Ki}\bfh^T_K(\bfD_{Ki};\bftheta_K))^T.$$
% The asymptotically equivalent version of weighted estimator can be obtained by solving $$\sum_{i=1}^n\hat{p}_ig(D_{0i}, \beta)=0$$ 
% add how to solve p
However, the dimension of stacking multiple over-identified functions could be substantially high when more secondary data are considered and those could be in different types and scales. Because of these multi-dimensional and complex data structure, the estimation algorithm might not converge (detected by our previous simulations), thus hindering the application of this method. Our MinBo estimator, on the other hand, is an effective solution making information integration feasible in practice when multiple secondary outcomes are available. Note that the semiparametric theory (if exist) will become much more complicated if we only assume a parametric form for $\bfD_{0i}$ and leave all others not specified, which merits substantial effort and can be pursued in future work.
% \red{actually I am not quite understand his question by not specifying any others. Double check the following form. Do you mean $X_{0i}$? What is $\eta(\cdot)$? If you write in this way, I think here $\theta$ will be different with the notation defined above, right?} 
% Then, a parametric sub-model conditional on covariates $\bfX_i$ should be in the form of 
% $$f(Y_{0i}|\bfX_{i};\bfbeta)f( \bfY_{1i}, \ldots, \bfY_{Ki}|\bfX_{i},\bfY_{0i};\bfeta(\bfbeta)),$$ with a nuisance parameter vector $\bfeta(\bfbeta)$. It can be seen that the nuisance parameter vector $\bftheta(\bfbeta)$ will be implicitly related to the main parameter $\bfbeta$ unless the secondary outcomes are conditional independent of the primary outcome. This will create difficulty to find the nuisance tangent space and thus merit substantial effort, which can definitely be pursued in future work but not the focus in this paper. 
}

\section{Extra simulation results}\label{Extra simulation results}

This section contains more detailed specifications and extra simulation results of the main paper. The covariates and residuals in first two secondary secondary models are generated as follows: time-dependent covariate vector $\tilde{\bfX}_{1i1}=\tilde{\bfX}_{2i1}$ follows multivariate Bernoulli distribution with success probability equal to $0.5$ and exchangeable correlation matrix with correlation coefficient equal to $0.3$; another time-dependent covariate vector $\tilde{\bfX}_{1i2}=\tilde{\bfX}_{2i2}$ follows multivariate Normal distribution with mean zero, variance one, and exchangeable correlation structure with correlation coefficient equal to $0.3$;  the last covariate vector $\tilde{\bfX}_{1i3}=\tilde{\bfX}_{2i3}$ is time-independent where all components are equal and are followed by Bernoulli distribution with success probability $0.5$. The residual vector in two data sets $\bfepsilon_i=(\bfepsilon_{1i}^T,\bfepsilon_{2i}^T)^T$ with $\bfepsilon_{ki}=(\epsilon_{ki1},\ldots,\epsilon_{ki4})^T$, for $k=1,2$, follows multivariate Normal distribution with mean zeros and variance-covariance 
\begin{equation}
\begin{split}
        \bfV=\left(\begin{array}{llll}
    \bfV_{11} & \bfV_{12}\\
    \bfV_{21} & \bfV_{22}
    \end{array}
    \right),
\end{split}
\end{equation}
where $\bfV_{11}$ and $\bfV_{22}$ are variance-covariance matrices for $\bfepsilon_{1i}$ and $\bfepsilon_{2i}$, respectively. The diagonal values of $\bfV_{11}$ are equal to one, and the off-diagonal values are all $0.8$; the diagonal values of $\bfV_{22}$ are equal to one, and the off-diagonal values are all $0.5$; the matrix $\bfV_{12}=\bfV_{21}^T$ has all elements equal to $\rho$, which controls the 
association strength between two longitudinal secondary outcomes.

Moreover, to construct the over-identified working estimating function, we adopt four basis matrices $\bfV_1$, $\bfV_2$, $\bfV_3$, and $\bfV_4$, %AQ: Please check phrasing "select the basis matrices... into the functions".
where $\bfV_1$ is the identity matrix, $\bfV_2$ is a matrix with $0$ on the diagonal and $1$ off the diagonal, $\bfV_3$ is a matrix with $1$ on the two main off-diagonals and $0$ elsewhere, and $\bfV_4$ is a matrix with $1$ at the left-top and right-bottom corners and $0$ otherwise. These matrices have been commonly used in previous studies \citep{qu2000improving,tang2011empirical,chen2021improving}.

\begin{figure}[h]\label{workflow}
    \centering
    \includegraphics[width=5in]{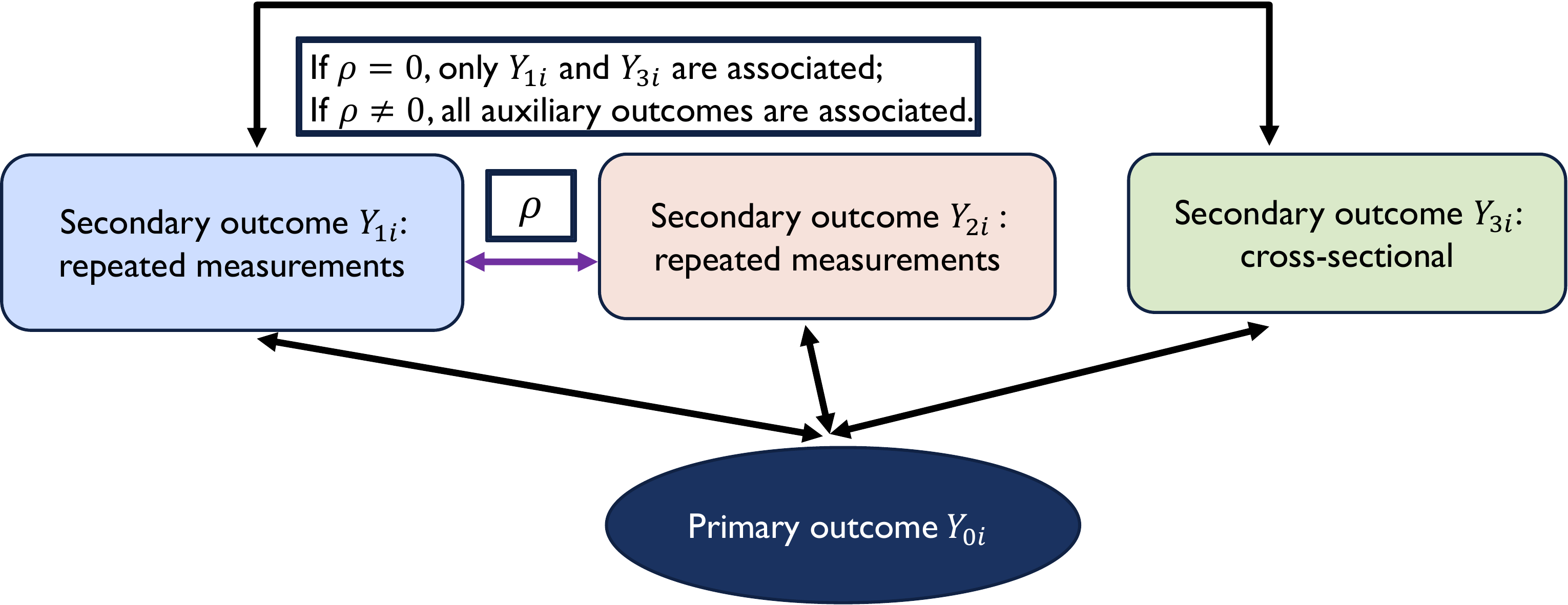}
    \caption{A schematic data structure \revision{for} simulation studies \revision{of} the main text of paper. The first and third secondary outcomes are always associated, and the first two secondary outcomes are associated controlled by $\rho$. All secondary outcomes are associated with primary outcome.}
    \label{simulation scheme}
\end{figure}

\revisiontwo{Figure \ref{workflow} provides a schematic data structure for simulation studies of the manuscript;} Table \ref{tableS1} summarizes the overall evaluations for eight estimation methods given correlation coefficient $\rho$ between two longitudinal auxiliary variable equal to $0.4$; Table \ref{tableS2} summarizes the results under $\rho=0$; Table \ref{tableS4} summarizes the results empirical relative efficiency (ERE) with partially observed secondary data; Table \ref{tableS3} summarizes overall evaluations under mis-specified mean structures where partial secondary data are observed. 
\revision{\subsection{Informative missingness}\label{Informative missingness}}
We also investigate the impact of informative missingness \citep{enders2010applied} from secondary data to the main parameter estimation. To generate partially observed secondary data with data missing not at random, we simulate informative observing indicator $R_{ki}$ for $k^{\text{th}}$ secondary data. The success probability of $R_{ki}$ is modeled as $\{1+\exp(-\bfX_{0i}^T\bfalpha)\}^{-1}$, for $k=1,2,3$, with $\bfalpha=(0.5,1,1,1)^T$ and $\bfX_{0i}$ defined in Section 3.1 \revision{of the main paper}. All other setups are the same to those in Section 3.1 of the main paper. To evaluate the performance of \revision{MinBo}, we only use mis-specified mean structures described in Section 3.2 of the main paper to construct three secondary models and only use secondary data with $R_{ki}=1$ to fit the model. Since observing indicators $R_{ki}$ depend on potential outcomes give mis-specified mean structures, it leads to the issue of informative missingness. The missing rates for each secondary data are around $25\%$, and the results are summarized in Table \ref{tableS5}. Given secondary data with informative missingness and mis-specified mean structures, the proposed estimates in the main model still have little estimation bias with satisfactory coverage probability close to $95\%$ nominal level. Moreover, in terms of ERE, \revision{MinBo} is still able to improve estimation efficiency under this context.\newline

\revision{\subsection{Little association between the primary and secondary outcomes}\label{Little association between the primary and secondary outcomes}
We consider a setup where the main data and three secondary data are generated in the same way as shown in Section 3.1 of the main paper, except that now the main outcome and three secondary outcomes are very mildly correlated (e.g., the correlation coefficient is around 0.15). We also consider different cases where the correlation coefficients $\rho$ between first two secondary outcomes are $0$, $0.4$, and $0.8$ , respectively. The results with the sample size of $600$ are summarized in Table \ref{little association} shown below.  From Table \ref{little association}, we find that the ERE values are barely above one for estimates based on the averaging scheme and the omnibus scheme. Note that the ERE values could be slightly lower than $1$ for estimates based on the aggregating scheme when $\rho$ is large. This is expected since the aggregating scheme may lower down the efficiency when secondary outcomes are highly associated with each other (Section 2.3 of the main paper). Further, note that our proposed method "MinBo" relies on the assumption that the main outcome and secondary outcomes are highly associated with each other. It implies that, if in practice these outcomes show very little association, there is indeed no need to integrate these secondary outcomes into the main analysis. In addition, even if the association strength is very mild, our method can still lead to comparable or a slightly better efficiency compared to that without considering secondary data. 
}

\revision{\subsection{Compare with the joint likelihood approach}\label{Compare with the joint likelihood approach}
To compare our method with the JL approach, we generate one primary outcome and two secondary outcomes based on multi-variate normal distribution, where the joint likelihood of three outcomes is easy to specify. Particularly, we consider cross-sectional primary outcome and two longitudinal secondary outcomes with four time points. The residuals for three outcomes are generated based on multi-variate normal distribution with mean zeros and variance ones. The correlation pattern among three outcomes are as follows: the correlation over time points within each secondary outcomes is $0.8$; the correlation between two secondary outcomes 
    is $0.6$; and the correlation between the primary outcome and secondary outcome is $0.8$. The covariates and effect sizes considered in the primary model and two secondary models are the same to those of Section 3.1 of the main paper. 
     
     We consider the averaging and aggregating schemes to borrow information from two secondary outcomes and compare with the MLE and the JL estimate. The JL approach is to jointly model three outcomes based on multivariate normal distribution with unstructured correlation pattern. To check the performance under mis-specified secondary models, we consider mean-structure mis-specification where the fitted means of two secondary outcomes only include two covariates (there are three covariates in total in the true mean-structure). The results are summarized in Table \ref{jl}. When the secondary models are correctly specified, we observe that both the JL estimate and MinBo estimate have little bias, and the JL estimate is more efficient than MinBo estimates in terms of ERE (ratio of MSE between the MLE and the target estimate, larger than one is more preferred, the larger the better). This is expected since the JL estimate is the most efficient one based on parametric theory \citep{newey1994large}. However, when the secondary models are mis-specified, the JL estimate leads to substantial bias and very small ERE (close to zero), whereas the MinBo estimates still have little bias and substantial efficiency gain in terms of ERE. This evaluation furthermore highlights the advantage of our proposed method, which is more preferred for practical use. 
     
     \revisiontwo{
     
     \subsection{The setting with higher correlations} \label{The setup with higher correlation}
     
     In order to assess the influence of the association strength between the main and secondary outcomes on efficiency gain, we generate another data that is the same to the one in the main paper, except that the conditional correlation coefficients between the main outcome and three secondary outcomes tend to be higher (about $0.7$ between the main outcome and two longitudinal secondary outcomes; around $0.6$ between the main outcome and the cross-sectional secondary outcome. In our setup from the main paper, these two correlation coefficients are only around $0.45$ and $0.3$, respectively). To be specific, we first generate a random variable $Z_{0i}$ following a standard normal distribution and then use this variable to generate the primary outcome $Y_{0i}=1$ if $Z_0\geq x_{0i}$ (with $x_{0i}$ defined in Section 3.1 of the main paper), and $Y_{0i}=0$ otherwise. Then, we generate two longitudinal secondary outcomes by $\bfY_{ki}=\bfX_{ki}\bftheta_{k}+\bfY_{0i}-\bfp_{0i}+0.4\bfepsilon_{ki}$, for $k=1,2$ (with $\bfX_{ki}$, $\bftheta_{k}$, $p_{0i}$, and $\bfepsilon_{ki}$ defined in Section 3.1 of the main paper. Here, the values in $\bfY_{0i}-\bfp_{0i}$ are the same across all observations for subject $i$), and the binary secondary outcome $Y_{3i}$ by $Y_{3i}=1$ if $-Z_0\geq x_{i}^\ast$, and $Y_{3i}=0$ otherwise (with $x_{i}^\ast$ defined in Section 3.1 of the main paper). All other setups are the same as the one in Section 3.1 of the main paper.  It can be seen from Table \ref{high cor} that the resulting MinBo estimators have significantly increased efficiency gain (from $1.2-1.3$ to $1.5-1.7$) compared to the results in the main paper (Table 2), even if all the models of secondary outcomes are mis-specified.
     
     }
}

\section{Extension to missing-data problem in the main model}\label{Extension to missing-data problem}
\revision{In the main paper}, we assume that all main data are observed or missing completely at random, which may be too ideal in some applications. However, by making a little modification, our method can be easily \revision{extended} to address the missing-data problem \revision{in the main model}. Hereby we present an extension by incorporating the technique, called inverse probability weight, a commonly used approach to deal with the issue where the main data are missing at random \citep{enders2010applied}. To be specific, let us first denote an observing indicator ${R}_{0i}$, which is equal to one if the main data from $i^{th}$ subject are observed and 0 otherwise. Then we define the probability of observing $i^{th}$ subject as $\pi_i(\bfalpha)=E(R_{0i}|\tilde{\bfX}_{0i})$, where $\tilde{\bfX}_{0i}$ and $\bfalpha$ are some covariates and parameters involved in the missing data model, respectively. Furthermore, let us define the score function for parameters $\bfalpha$ in the missing data model as $\sum_{i=1}^n\bfL_i(\bfalpha)$. We keep other setups the same to Section 2 \revision{of} the main paper. Then, we simultaneously solve $\bfbeta$ and $\bfalpha$ by the following modified weighted estimating equations:
\begin{equation}\label{missing data}
    \sum_{i=1}^{n}{p}^\ast_i\tilde{\bfg}(\bfD_{0i},{\bfalpha};\bfbeta)=\bfzero,
\end{equation}
where \revision{weights} ${p}^\ast_i$ are re-constructed from our proposed schemes (Section 2.2, 2.3, and 2.4 \revision{of} the main paper); $\tilde{\bfg}(\bfD_{0i},{\bfalpha};\bfbeta)=(R_{0i}/\pi_i({\bfalpha})\bfg^T(\bfD_{0i};\bfbeta),\bfL^T_i({\bfalpha}))^T$ with $\bfg(\bfD_{0i};\bfbeta)$ the same to (1) of the main paper. Asymptotic properties then can be similarly derived by following the lines of proofs in previous two sections. In the presence of high missingness, we can see that the only modification is to construct an extended estimating functions $\tilde{\bfg}(\bfD_{0i},{\bfalpha};\bfbeta)$ for the main analysis. The first component in $\tilde{\bfg}(\bfD_{0i},{\bfalpha};\bfbeta)$ is a typical estimating function under inverse probability weight framework. The second \revision{component} is to make the constructed estimating functions $\tilde{\bfg}(\bfD_{0i},{\bfalpha};\bfbeta)$ independently and identically distributed, in order to theoretically guarantee the efficiency gain for the estimation of $\bfbeta$. Evaluations for this extension is not the primary focus in this article and thus is regarded as future work.

\revision{
\section{Over-identified estimating functions}\label{Over-identified estimating functions}

This section provides examples of the over-identified estimation functions for fitting the secondary data. There are two commonly used over-identified estimating functions $\bfh_{k}(\bfD_{ki};\bftheta_k)$ from the literature. One is for repeated secondary outcomes, i.e., $\bfY_{ki}$ are repeated measurements for each subject during the follow-up:
\begin{equation}\label{repeated measurements}
	\begin{aligned}
	    \bfh_k(\bfD_{ki};\bftheta_k)=\left(\begin{array}{l}
    \bfZ_{ki}^T \tilde{\bfR}_{ki}^{-1/2}\bfV_{1_k}\tilde{\bfR}_{ki}^{-1/2}\big\{{\bfY}_{ki}-\bfmu_k({\bfX}_{ki};\bftheta_k)\big\}\\
    \ldots\\
    \bfZ_{ki}^T \tilde{\bfR}_{ki}^{-1/2}\bfV_{\tau_k}\tilde{\bfR}_{ki}^{-1/2}\big\{{\bfY}_{ki}-\bfmu_k({\bfX}_{ki};\bftheta_k)\big\}
    \end{array}
    \right), 
	\end{aligned}
    \end{equation}	
where $\bfZ_{ki}=\partial\bfmu_{k}({\bfX}_{ki};\bftheta_k)/\partial\bftheta_k^T$; $\tilde{\bfR}_{ki}$ is a diagonal matrix containing the variance of each element in ${\bfY}_{ki}$; $\bfmu_k({\bfX}_{ki};\bftheta_k)$ are the conditional means of the $k^{\text{th}}$ secondary outcome ${\bfY}_{ki}$ indexed by parameters $\bftheta_k$. The $\tau_k$ matrices $\bfV_j$ lead to the over-identified estimating function with length $r_k\times\tau_k$, given the dimensions of $\bftheta_k$, $r_k$, and $\tau_k\geq 2$. $\bfV_{1_k},\ldots,\bfV_{\tau_k}$ can be a sequence of base matrices \citep{qu2000improving,tang2011empirical,chen2021improving}. The other form of $\bfh_{k}(\bfD_{ki};\bftheta_k)$ is designed for cross-sectional secondary outcome $Y_{ki}$, i.e.,
\begin{equation}\label{surrogate}
    \bfh_k(\bfD_{ki};\bftheta_k)=\bfd({\bfX}_{ki},\tilde{\bfZ}_{ki};\bftheta_k)\big\{{Y}_{ki}-\mu_k({\bfX}_{ki};\bftheta_k)\big\},
\end{equation}
where $\bfd({\bfX}_{ki},\tilde{\bfZ}_{ki};\bftheta_k)$ is a user-specified vector function with dimension larger than that of $\bftheta_k$, which typically takes the form of $\bfd({\bfX}_{ki},\tilde{\bfZ}_{ki};\bftheta_k)=({\bfX}^T_{ki},\tilde{\bfZ}^T_{ki})^T$. Note that the vector $\tilde{\bfZ}_{ki}$ contains the redundant variables satisfying $E\big\{\bfh_k(\bfD_{ki};\bftheta^\ast_k)\big\}=\bfzero$, for some parameter values $\bftheta^\ast_k$. Both working functions in (\ref{repeated measurements}) and (\ref{surrogate}) could be applied to continuous and categorical outcomes.
}

\revision{
\section{\revisiontwo{More discussion on the estimation efficiency}}\label{Discussion based on Semi-parametric theory}
\revisiontwo{Based on the discussion in Section \ref{Other forms}, we have shown that the proposed estimation procedure (2) and (3) of the main paper is semi-parametric efficient given the information of $E\{\bfg(\bfD_{0i};\bfbeta)\}=\bfzero$ and $E\{R_{ki}\bfh_k(\bfD_{ki};\bftheta_k)\}=\bfzero$. In this section, we provide more discussion on estimation efficiency of this estimator.} 
    
First of all, the proposed estimate will still be consistent and does not violate the semi-parametric theory. Note that based on the theory of M-estimate \citep{newey1994large}, the estimation consistency for a semi-parametric estimate is guaranteed by $E[\bfW(\bfX_0)\{Y_0-\mu(\bfX_0; \bfbeta)\}]=\bfzero$ ($\bfW(\bfX_0)$ is some weight function depending on the covariates $\bfX_0$), whereas our proposed estimate is based on the first moment $E\{p_\ast \bfg(\bfD_0;\bfbeta)\}=\bfzero$, where $p_\ast$ is the limiting value of $\hat{p}_i$, and $\bfg(\bfD_{0i};\bfbeta)$ is some valid estimating function. Based on the empirical likelihood theory, we have $n\hat{p}_i=1+o_p(1)$ \citep{qin1994empirical,owen2001empirical}, which implies that $np_\ast=1$. Thus, in terms of the first moment that the estimation consistency relies on, our proposed estimate does not introduce the dependency between the main outcome and the secondary outcome, and the proposed estimate is consistent by realizing $E\{np_\ast \bfg(\bfD_0;\bfbeta)\}=E\{ \bfg(\bfD_0;\bfbeta)\}=\bfzero$. 
    
% One quick note is that we require $g(D_{0i};\theta)$ be a valid estimating function for $\theta$, which could be in some form of $W(X_{0i})\{Y_{0i}-p(X_{0i})\}$. The function $Y_{0i}-p(X_{0i})$ alone may not be a valid estimating function for $\theta$ since it does not set any association constraint between the covariate $X_{0i}$ and the residual $Y_{0i}-p(X_{0i})$. \red{just want to double check this terminology of "valid". I think $W(x)$ could be identity, right?}
    
Second, the proposed estimate could be more efficient than the semi-parametric estimates without considering any secondary data. In theory, we allow a general form of $\bfg(\bfD_{0i};\bfbeta)$, which could be any valid estimating function for $\bfbeta$. In our paper, we focus on binary outcomes from cross-sectional studies, thus we consider the form of estimating function $\bfg(\bfD_{0i};\bfbeta)=\bfX_{0i}\{Y_{0i}-\mu(\bfX_{0i};\bfbeta)\}$ by considering a logit link function for numerical studies (simulation and real data application). Note that this is the score function from logistic regression and is the estimating function with semi-parametric efficiency (indeed this function will lead to maximum likelihood estimate, which is the most efficient estimate). Under this context, by considering the information from the secondary outcomes through the estimating equation $\sum_{i=1}^n\hat{p}_i\bfg(\bfD_{0i},\bfbeta)=\bfzero$, we have shown in both theory and simulation that the proposed estimate is more efficient than the estimate without information integration. This implies that the proposed estimate could be more efficient than the semi-parametric estimate solved by $\sum_{i=1}^n[\bfW(\bfX_{0i})\{Y_{0i}-\mu(\bfX_{0i}; \bfbeta)\}]=\bfzero$ without considering any secondary data, which has been shown under the scenario with cross-sectional binary outcomes.
    
Finally, the above discussion can be extended to a more general framework as follows. Let's consider the data $Y_{0i}$, $\bfY_{si}$, and $\bfX_i$, where $Y_{0i}$ is the main outcome, $\bfY_{si}$ are (multiple) secondary outcome(s), and $\bfX_i$ is the vector containing all covariates in the main and secondary data. Thereafter, we define the most efficient estimate ($E_1$), which is obtained by maximizing the joint likelihood $L(Y_{0i},\bfY_{si},\bfX_i)$ (suppose it is correctly specified), the semi-parametric estimate ($E_2$) by considering the data $Y_{0i}$ and $\bfX_i$, and our proposed estimate ($E_3$) by using the semi-parametric estimating function with weights $\hat{p}_i$ for information borrowing. Thus, the estimation efficiencies of these three estimates are ordered as $E_1\geq E_3 \geq E_2$. It means our proposed estimate is a compromised estimate between the most efficient estimate and the semi-parametric efficient estimate, by effectively integrating information from (multiple) secondary outcomes. Note that the full likelihood estimate ($E_1$) may not be accessible in many applications because it requires correct specification of the full likelihood, which is difficult to achieve in practice (Table \ref{jl}). Of note is that the main outcome and (multiple) secondary outcomes could follow different distributions (e.g., continuous/count/binary) and even in different data structures (e.g., cross-sectional, longitudinal), which introduce challenges for distribution specification. Our proposed estimate provides a convenient and robust way to leverage information from secondary outcomes for improving primary parameters efficiency.
}

\bibliographystyle{plainnat} 
\bibliography{biomsample_bib}

\begin{table}
\caption{\label{tableS1}Overall evaluations for eight estimation methods (defined in the simulation section \revision{of} the main paper) under the combination of different sample sizes ($n$) and proportions for observing three secondary data (CASE 1: $\eta_1=\eta_2=\eta_3=1$; CASE 2: $\eta_1=0.6$,$\eta_2=0.7$,$\eta_3=0.5$), given the correlation ($\rho$) between two longitudinal secondary outcome equal to $0.4$. MCSE: Monte Carlo standard error; ASE: asymptotic standard error; CP: coverage probability. All values have been  multiplied by $100$.}
\centering
\tiny
\begin{tabular}{cccccccccccc}
\toprule
       &           &          & \multicolumn{4}{c}{n=300}  &  & \multicolumn{4}{c}{n=600}  \\\cline{4-7} \cline{9-12} 
       &           &          & Bias & MCSE & ASE & 95\%CP &  & Bias & MCSE & ASE & 95\%CP \\\midrule
CASE 1 & single100 & $\beta_0$ & 2.6  & 24   & 24  & 95     &  & 0.8  & 17   & 17  & 94     \\
       &           & $\beta_1$ & -0.5 & 26   & 25  & 93     &  & 0.1  & 18   & 18  & 94     \\
       &           & $\beta_2$ & -2.6 & 15   & 15  & 95     &  & -1.2 & 11   & 11  & 94     \\
       &           & $\beta_3$ & -0.8 & 29   & 28  & 95     &  & 0.1  & 20   & 20  & 95     \\
       & single010 & $\beta_0$ & 2.2  & 24   & 24  & 95     &  & 1.0  & 17   & 17  & 95     \\
       &           & $\beta_1$ & -0.3 & 28   & 28  & 95     &  & -0.3 & 20   & 20  & 95     \\
       &           & $\beta_2$ & -2.4 & 15   & 15  & 96     &  & -0.9 & 11   & 11  & 95     \\
       &           & $\beta_3$ & -0.6 & 25   & 25  & 96     &  & 0.2  & 18   & 18  & 95     \\
       & single001 & $\beta_0$ & 2.7  & 24   & 24  & 95     &  & 0.9  & 17   & 17  & 94     \\
       &           & $\beta_1$ & -0.9 & 26   & 26  & 94     &  & -0.3 & 19   & 18  & 94     \\
       &           & $\beta_2$ & -2.6 & 16   & 15  & 95     &  & -1.1 & 11   & 11  & 94     \\
       &           & $\beta_3$ & -0.7 & 28   & 28  & 95     &  & 0.1  & 20   & 19  & 96     \\
       & ave101    & $\beta_0$ & 2.6  & 24   & 23  & 95     &  & 0.8  & 17   & 16  & 95     \\
       &           & $\beta_1$ & -0.7 & 25   & 24  & 94     &  & -0.1 & 18   & 17  & 94     \\
       &           & $\beta_2$ & -2.5 & 15   & 15  & 95     &  & -1.1 & 11   & 10  & 94     \\
       &           & $\beta_3$ & -0.7 & 28   & 27  & 95     &  & 0.1  & 20   & 19  & 95     \\
       & agg101    & $\beta_0$ & 2.6  & 25   & 23  & 94     &  & 0.6  & 17   & 16  & 94     \\
       &           & $\beta_1$ & -0.9 & 26   & 25  & 93     &  & 0.2  & 18   & 17  & 94     \\
       &           & $\beta_2$ & -2.3 & 16   & 15  & 93     &  & -1.0 & 11   & 11  & 94     \\
       &           & $\beta_3$ & -0.8 & 29   & 27  & 94     &  & 0.0  & 20   & 19  & 95     \\
       & ave111    & $\beta_0$ & 2.4  & 24   & 23  & 95     &  & 0.9  & 17   & 16  & 95     \\
       &           & $\beta_1$ & -0.5 & 26   & 25  & 94     &  & -0.1 & 18   & 18  & 94     \\
       &           & $\beta_2$ & -2.4 & 15   & 14  & 95     &  & -1.0 & 11   & 10  & 94     \\
       &           & $\beta_3$ & -0.7 & 27   & 26  & 95     &  & 0.1  & 19   & 18  & 95     \\
       & agg111    & $\beta_0$ & 2.4  & 24   & 22  & 94     &  & 0.5  & 17   & 16  & 95     \\
       &           & $\beta_1$ & -0.8 & 26   & 25  & 93     &  & 0.2  & 18   & 17  & 94     \\
       &           & $\beta_2$ & -2.0 & 17   & 16  & 94     &  & -0.8 & 12   & 11  & 94     \\
       &           & $\beta_3$ & -0.7 & 26   & 25  & 94     &  & 0.1  & 18   & 17  & 94     \\
       & omn111    & $\beta_0$ & 2.4  & 24   & 22  & 95     &  & 0.7  & 16   & 16  & 95     \\
       &           & $\beta_1$ & -0.8 & 25   & 24  & 93     &  & 0.0  & 18   & 17  & 94     \\
       &           & $\beta_2$ & -2.2 & 15   & 14  & 94     &  & -0.9 & 11   & 10  & 93     \\
       &           & $\beta_3$ & -0.6 & 26   & 25  & 95     &  & 0.1  & 18   & 18  & 95     \\\toprule
       &           &          & \multicolumn{4}{c}{n=300}  &  & \multicolumn{4}{c}{n=600}  \\\cline{4-7} \cline{9-12} 
       &           &          & Bias & MCSE & ASE & 95\%CP &  & Bias & MCSE & ASE & 95\%CP \\\midrule
CASE 2 & single100 & $\beta_0$ & 2.7  & 25   & 24  & 95     &  & 1.0  & 17   & 17  & 95     \\
       &           & $\beta_1$ & -0.4 & 27   & 26  & 95     &  & -0.2 & 19   & 18  & 94     \\
       &           & $\beta_2$ & -2.9 & 16   & 16  & 95     &  & -1.3 & 12   & 11  & 94     \\
       &           & $\beta_3$ & -0.8 & 29   & 28  & 95     &  & 0.2  & 20   & 20  & 94     \\
       & single010 & $\beta_0$ & 2.6  & 24   & 24  & 95     &  & 0.9  & 17   & 17  & 95     \\
       &           & $\beta_1$ & -0.3 & 28   & 28  & 95     &  & -0.3 & 20   & 20  & 95     \\
       &           & $\beta_2$ & -2.6 & 16   & 16  & 95     &  & -1.2 & 12   & 11  & 94     \\
       &           & $\beta_3$ & -1.2 & 27   & 27  & 96     &  & 0.3  & 19   & 19  & 95     \\
       & single001 & $\beta_0$ & 2.6  & 25   & 24  & 95     &  & 1.0  & 17   & 17  & 95     \\
       &           & $\beta_1$ & -0.7 & 27   & 26  & 94     &  & -0.4 & 19   & 19  & 94     \\
       &           & $\beta_2$ & -3.0 & 16   & 16  & 94     &  & -1.1 & 11   & 11  & 94     \\
       &           & $\beta_3$ & -0.5 & 29   & 28  & 95     &  & 0.2  & 20   & 19  & 96     \\
       & ave101    & $\beta_0$ & 2.6  & 24   & 24  & 95     &  & 1.0  & 17   & 17  & 96     \\
       &           & $\beta_1$ & -0.6 & 27   & 26  & 95     &  & -0.3 & 19   & 18  & 94     \\
       &           & $\beta_2$ & -2.9 & 16   & 15  & 95     &  & -1.2 & 11   & 11  & 94     \\
       &           & $\beta_3$ & -0.7 & 28   & 27  & 95     &  & 0.2  & 20   & 19  & 96     \\
       & agg101    & $\beta_0$ & 2.6  & 25   & 24  & 94     &  & 0.8  & 17   & 17  & 95     \\
       &           & $\beta_1$ & -0.7 & 27   & 26  & 94     &  & -0.3 & 19   & 18  & 94     \\
       &           & $\beta_2$ & -3.0 & 16   & 15  & 93     &  & -1.1 & 12   & 11  & 93     \\
       &           & $\beta_3$ & -0.6 & 29   & 27  & 94     &  & 0.2  & 20   & 19  & 95     \\
       & ave111    & $\beta_0$ & 2.6  & 24   & 23  & 95     &  & 1.0  & 17   & 17  & 95     \\
       &           & $\beta_1$ & -0.5 & 27   & 26  & 95     &  & -0.3 & 19   & 18  & 95     \\
       &           & $\beta_2$ & -2.8 & 15   & 15  & 95     &  & -1.2 & 11   & 11  & 94     \\
       &           & $\beta_3$ & -0.7 & 28   & 27  & 95     &  & 0.2  & 19   & 19  & 96     \\
       & agg111    & $\beta_0$ & 2.8  & 25   & 23  & 94     &  & 0.7  & 17   & 16  & 95     \\
       &           & $\beta_1$ & -0.5 & 27   & 26  & 93     &  & -0.3 & 19   & 18  & 94     \\
       &           & $\beta_2$ & -2.8 & 17   & 16  & 94     &  & -1.1 & 12   & 11  & 93     \\
       &           & $\beta_3$ & -1.1 & 28   & 26  & 94     &  & 0.3  & 19   & 18  & 94     \\
       & omn111    & $\beta_0$ & 2.6  & 25   & 23  & 94     &  & 0.8  & 17   & 16  & 95     \\
       &           & $\beta_1$ & -0.6 & 27   & 25  & 94     &  & -0.4 & 19   & 18  & 94     \\
       &           & $\beta_2$ & -2.9 & 16   & 15  & 94     &  & -1.1 & 11   & 11  & 93     \\
       &           & $\beta_3$ & -0.7 & 28   & 26  & 94     &  & 0.3  & 19   & 19  & 95  \\\bottomrule  
\end{tabular}
\end{table}

\begin{table}
\caption{\label{tableS2}Overall evaluations for eight estimation methods (defined in the simulation section \revision{of} the main paper) under the combination of different sample sizes ($n$) and proportions for observing three secondary data (CASE 1: $\eta_1=\eta_2=\eta_3=1$; CASE 2: $\eta_1=0.6$,$\eta_2=0.7$,$\eta_3=0.5$), given the correlation ($\rho$) between two longitudinal secondary outcome equal to $0$. MCSE: Monte Carlo standard error; ASE: asymptotic standard error; CP: coverage probability. All values have been  multiplied by $100$.}
\centering
\tiny
\begin{tabular}{lccccccccccc}
\toprule
       &           &          & \multicolumn{4}{c}{n=300}  &  & \multicolumn{4}{c}{n=600}  \\ \cline{4-7} \cline{9-12} 
       &           &          & Bias & MCSE & ASE & 95\%CP &  & Bias & MCSE & ASE & 95\%CP \\\midrule
CASE 1 & single100 & $\beta_0$ & 2.7  & 25   & 24  & 95     &  & 1.2  & 17   & 17  & 95     \\
       &           & $\beta_1$ & -0.7 & 27   & 26  & 94     &  & -0.4 & 19   & 18  & 95     \\
       &           & $\beta_2$ & -2.8 & 16   & 15  & 94     &  & -1.2 & 11   & 11  & 94     \\
       &           & $\beta_3$ & -0.6 & 29   & 28  & 95     &  & 0.1  & 20   & 20  & 95     \\
       & single010 & $\beta_0$ & 2.6  & 25   & 24  & 95     &  & 1.4  & 17   & 17  & 96     \\
       &           & $\beta_1$ & -0.8 & 28   & 28  & 95     &  & -0.7 & 20   & 20  & 95     \\
       &           & $\beta_2$ & -2.7 & 15   & 16  & 95     &  & -1.1 & 11   & 11  & 95     \\
       &           & $\beta_3$ & -0.6 & 26   & 26  & 96     &  & 0.0  & 18   & 18  & 95     \\
       & single001 & $\beta_0$ & 2.7  & 25   & 24  & 95     &  & 1.3  & 17   & 17  & 95     \\
       &           & $\beta_1$ & -1.1 & 27   & 27  & 95     &  & -0.7 & 19   & 19  & 95     \\
       &           & $\beta_2$ & -2.6 & 16   & 16  & 94     &  & -1.1 & 11   & 11  & 94     \\
       &           & $\beta_3$ & -0.5 & 28   & 27  & 95     &  & 0.0  & 20   & 19  & 95     \\
       & ave101    & $\beta_0$ & 2.7  & 24   & 23  & 94     &  & 1.2  & 17   & 16  & 95     \\
       &           & $\beta_1$ & -0.9 & 26   & 25  & 95     &  & -0.6 & 18   & 18  & 95     \\
       &           & $\beta_2$ & -2.6 & 15   & 15  & 94     &  & -1.1 & 11   & 11  & 94     \\
       &           & $\beta_3$ & -0.5 & 28   & 27  & 95     &  & 0.1  & 20   & 19  & 95     \\
       & agg101    & $\beta_0$ & 2.7  & 25   & 23  & 94     &  & 1.1  & 17   & 16  & 95     \\
       &           & $\beta_1$ & -1.0 & 26   & 24  & 94     &  & -0.4 & 18   & 17  & 94     \\
       &           & $\beta_2$ & -2.6 & 16   & 15  & 93     &  & -1.1 & 11   & 10  & 94     \\
       &           & $\beta_3$ & -0.5 & 29   & 27  & 94     &  & 0.0  & 20   & 19  & 95     \\
       & ave111    & $\beta_0$ & 2.6  & 24   & 23  & 94     &  & 1.2  & 17   & 16  & 95     \\
       &           & $\beta_1$ & -0.9 & 27   & 26  & 94     &  & -0.6 & 19   & 18  & 95     \\
       &           & $\beta_2$ & -2.6 & 15   & 15  & 94     &  & -1.1 & 11   & 11  & 94     \\
       &           & $\beta_3$ & -0.5 & 27   & 26  & 95     &  & 0.1  & 19   & 19  & 95     \\
       & agg111    & $\beta_0$ & 2.7  & 24   & 22  & 95     &  & 1.1  & 16   & 16  & 95     \\
       &           & $\beta_1$ & -1.0 & 26   & 24  & 93     &  & -0.4 & 18   & 17  & 94     \\
       &           & $\beta_2$ & -2.6 & 16   & 15  & 94     &  & -1.0 & 11   & 10  & 94     \\
       &           & $\beta_3$ & -0.6 & 26   & 25  & 94     &  & -0.1 & 18   & 18  & 94     \\
       & omn111    & $\beta_0$ & 2.6  & 24   & 23  & 94     &  & 1.2  & 16   & 16  & 95     \\
       &           & $\beta_1$ & -1.0 & 26   & 25  & 93     &  & -0.6 & 18   & 18  & 95     \\
       &           & $\beta_2$ & -2.5 & 15   & 14  & 94     &  & -1.0 & 11   & 10  & 94     \\
       &           & $\beta_3$ & -0.5 & 27   & 26  & 95     &  & 0.0  & 19   & 18  & 95     \\\toprule
       &           &          & \multicolumn{4}{c}{n=300}  &  & \multicolumn{4}{c}{n=600}  \\ \cline{4-7} \cline{9-12} 
       &           &          & Bias & MCSE & ASE & 95\%CP &  & Bias & MCSE & ASE & 95\%CP \\\midrule
CASE 2 & single100 & $\beta_0$ & 2.9  & 25   & 24  & 94     &  & 1.4  & 17   & 17  & 95     \\
       &           & $\beta_1$ & -0.7 & 28   & 27  & 94     &  & -0.7 & 19   & 19  & 95     \\
       &           & $\beta_2$ & -2.9 & 16   & 16  & 95     &  & -1.3 & 12   & 11  & 94     \\
       &           & $\beta_3$ & -0.7 & 29   & 28  & 96     &  & 0.2  & 20   & 20  & 95     \\
       & single010 & $\beta_0$ & 2.8  & 25   & 24  & 95     &  & 1.3  & 17   & 17  & 95     \\
       &           & $\beta_1$ & -0.7 & 28   & 28  & 95     &  & -0.7 & 20   & 20  & 95     \\
       &           & $\beta_2$ & -2.8 & 16   & 16  & 95     &  & -1.2 & 11   & 11  & 96     \\
       &           & $\beta_3$ & -1.0 & 27   & 27  & 96     &  & 0.3  & 19   & 19  & 95     \\
       & single001 & $\beta_0$ & 2.8  & 25   & 24  & 95     &  & 1.3  & 17   & 17  & 95     \\
       &           & $\beta_1$ & -1.1 & 28   & 27  & 95     &  & -0.9 & 19   & 19  & 95     \\
       &           & $\beta_2$ & -3.0 & 16   & 16  & 94     &  & -1.1 & 11   & 11  & 94     \\
       &           & $\beta_3$ & -0.3 & 28   & 28  & 95     &  & 0.2  & 20   & 19  & 95     \\
       & ave101    & $\beta_0$ & 2.8  & 25   & 24  & 95     &  & 1.3  & 17   & 17  & 95     \\
       &           & $\beta_1$ & -0.9 & 27   & 26  & 95     &  & -0.8 & 19   & 18  & 94     \\
       &           & $\beta_2$ & -2.9 & 16   & 16  & 94     &  & -1.2 & 11   & 11  & 94     \\
       &           & $\beta_3$ & -0.5 & 28   & 27  & 95     &  & 0.2  & 20   & 19  & 95     \\
       & agg101    & $\beta_0$ & 2.9  & 25   & 23  & 94     &  & 1.3  & 17   & 17  & 95     \\
       &           & $\beta_1$ & -1.0 & 27   & 26  & 94     &  & -0.8 & 19   & 18  & 94     \\
       &           & $\beta_2$ & -3.1 & 16   & 15  & 93     &  & -1.2 & 11   & 11  & 94     \\
       &           & $\beta_3$ & -0.4 & 29   & 27  & 95     &  & 0.2  & 20   & 19  & 95     \\
       & ave111    & $\beta_0$ & 2.8  & 25   & 24  & 95     &  & 1.3  & 17   & 17  & 95     \\
       &           & $\beta_1$ & -0.9 & 27   & 26  & 95     &  & -0.8 & 19   & 19  & 95     \\
       &           & $\beta_2$ & -2.9 & 16   & 15  & 95     &  & -1.2 & 11   & 11  & 94     \\
       &           & $\beta_3$ & -0.6 & 28   & 27  & 96     &  & 0.2  & 19   & 19  & 95     \\
       & agg111    & $\beta_0$ & 3.1  & 25   & 23  & 94     &  & 1.2  & 17   & 16  & 95     \\
       &           & $\beta_1$ & -0.9 & 27   & 26  & 94     &  & -0.8 & 19   & 18  & 94     \\
       &           & $\beta_2$ & -3.2 & 17   & 15  & 93     &  & -1.3 & 11   & 11  & 95     \\
       &           & $\beta_3$ & -0.9 & 28   & 26  & 95     &  & 0.3  & 19   & 19  & 95     \\
       & omn111    & $\beta_0$ & 2.9  & 25   & 23  & 94     &  & 1.2  & 17   & 16  & 96     \\
       &           & $\beta_1$ & -1.0 & 27   & 26  & 94     &  & -0.9 & 19   & 18  & 94     \\
       &           & $\beta_2$ & -3.1 & 16   & 15  & 93     &  & -1.2 & 11   & 11  & 94     \\
       &           & $\beta_3$ & -0.5 & 28   & 27  & 95     &  & 0.3  & 19   & 19  & 95   \\\bottomrule 
\end{tabular}
\end{table}

\begin{table}
\caption{Empirical relative efficiency for eight estimators (defined in the simulation section \revision{of} the main paper) under the combination of different sample sizes ($n$) and correlation between two longitudinal secondary outcomes ($\rho$). All secondary data are partially observed ($\eta_1=0.6$,$\eta_2=0.7$,$\eta_3=0.5$).}
\label{tableS4}
\scriptsize
\centering
\begin{tabular}{lccccccccccc}
\toprule
         &         &         &  & \multicolumn{8}{c}{Empirical relative efficiency}                                     \\\cline{5-12} 
         &         &         &  & single100 & single010 & single001 & ave110 & agg110 & ave111 & agg111 & omn111 \\\midrule
n=300  & $\rho$=0   & $\beta_0$ &  & 0.99      & 1.02      & 1.00      & 1.02   & 0.98   & 1.03   & 1.00   & 1.02   \\
    &         & $\beta_1$ &  & 1.04      & 1.00      & 1.06      & 1.11   & 1.11   & 1.09   & 1.10   & 1.12   \\
         &         & $\beta_2$ &  & 0.99      & 1.03      & 0.99      & 1.03   & 0.97   & 1.05   & 0.95   & 1.02   \\
         &         & $\beta_3$ &  & 0.97      & 1.10      & 1.01      & 1.01   & 0.97   & 1.05   & 1.05   & 1.04   \\
         & $\rho$=0.4 & $\beta_0$ &  & 1.00      & 1.03      & 1.00      & 1.03   & 0.97   & 1.05   & 1.00   & 1.02   \\
         &         & $\beta_1$ &  & 1.08      & 1.00      & 1.11      & 1.15   & 1.09   & 1.14   & 1.08   & 1.14   \\
         &         & $\beta_2$ &  & 1.04      & 1.06      & 1.02      & 1.07   & 0.97   & 1.09   & 0.91   & 1.03   \\
         &         & $\beta_3$ &  & 0.97      & 1.13      & 1.00      & 1.00   & 0.96   & 1.05   & 1.06   & 1.05   \\
         & $\rho$=0.8 & $\beta_0$ &  & 1.00      & 1.03      & 1.00      & 1.02   & 0.93   & 1.05   & 0.95   & 1.02   \\
         &         & $\beta_1$ &  & 1.09      & 1.00      & 1.15      & 1.16   & 0.98   & 1.15   & 0.97   & 1.14   \\
         &         & $\beta_2$ &  & 1.03      & 1.08      & 1.05      & 1.07   & 0.87   & 1.10   & 0.77   & 1.03   \\
         &         & $\beta_3$ &  & 0.97      & 1.16      & 0.99      & 0.99   & 0.94   & 1.05   & 1.06   & 1.05   \\\midrule
n=600  & $\rho$=0   & $\beta_0$ &  & 1.03      & 1.02      & 1.04      & 1.06   & 1.06   & 1.06   & 1.08   & 1.08   \\
    &         & $\beta_1$ &  & 1.07      & 1.01      & 1.04      & 1.10   & 1.11   & 1.09   & 1.12   & 1.11   \\
         &         & $\beta_2$ &  & 1.01      & 1.04      & 1.05      & 1.06   & 1.06   & 1.07   & 1.04   & 1.09   \\
         &         & $\beta_3$ &  & 0.99      & 1.10      & 1.02      & 1.02   & 1.01   & 1.06   & 1.10   & 1.08   \\
         & $\rho$=0.4 & $\beta_0$ &  & 1.04      & 1.02      & 1.04      & 1.06   & 1.04   & 1.07   & 1.07   & 1.08   \\
         &         & $\beta_1$ &  & 1.11      & 1.00      & 1.09      & 1.15   & 1.10   & 1.13   & 1.10   & 1.13   \\
         &         & $\beta_2$ &  & 1.04      & 1.05      & 1.08      & 1.09   & 1.04   & 1.10   & 0.97   & 1.09   \\
         &         & $\beta_3$ &  & 0.99      & 1.13      & 1.01      & 1.01   & 1.00   & 1.07   & 1.11   & 1.09   \\
         & $\rho$=0.8 & $\beta_0$ &  & 1.04      & 1.05      & 1.05      & 1.06   & 1.00   & 1.08   & 1.06   & 1.08   \\
         &         & $\beta_1$ &  & 1.12      & 1.00      & 1.15      & 1.17   & 1.00   & 1.15   & 0.99   & 1.10   \\
         &         & $\beta_2$ &  & 1.07      & 1.07      & 1.11      & 1.11   & 0.97   & 1.12   & 0.86   & 1.05   \\
         &         & $\beta_3$ &  & 0.99      & 1.16      & 1.01      & 1.01   & 0.99   & 1.07   & 1.14   & 1.10   \\\bottomrule
\end{tabular}
\end{table}

\begin{table}
\caption{ \label{tableS3}Overall evaluations for mis-specified working models under the combination of different sample sizes ($n$) and correlations ($\rho$), given three secondary data are partially observed ($\eta_1=0.6$,$\eta_2=0.7$,$\eta_3=0.5$). MCSE: Monte Carlo standard error; ASE: asymptotic standard error; ERE: empirical relative efficiency; CP: coverage probability. All values (except ERE) are multiplied by $100$.}
\scriptsize
\centering
\begin{tabular}{cccccccccccccc}
\toprule
        &        &          & \multicolumn{5}{c}{n=300}         &  & \multicolumn{5}{c}{n=600}         \\ \cline{4-8} \cline{10-14} 
        &        &          & Bias & MCSE & ASE & ERE  & 95\%CP &  & Bias & MCSE & ASE & ERE  & 95\%CP \\\midrule
$\rho$=0   & ave111 & $\beta_0$ & 2.7  & 25   & 24  & 1.03 & 95     &  & 1.3  & 17   & 17  & 1.04 & 95     \\
        &        & $\beta_1$ & -0.8 & 28   & 27  & 1.06 & 95     &  & -0.7 & 19   & 19  & 1.06 & 95     \\
        &        & $\beta_2$ & -2.9 & 16   & 16  & 1.01 & 94     &  & -1.2 & 11   & 11  & 1.04 & 94     \\
        &        & $\beta_3$ & -0.5 & 28   & 27  & 1.04 & 95     &  & 0.2  & 19   & 19  & 1.06 & 96     \\
        & ave101 & $\beta_0$ & 2.7  & 25   & 24  & 1.02 & 95     &  & 1.3  & 17   & 17  & 1.04 & 96     \\
        &        & $\beta_1$ & -0.8 & 28   & 27  & 1.07 & 95     &  & -0.7 & 19   & 19  & 1.08 & 95     \\
        &        & $\beta_2$ & -2.9 & 16   & 16  & 1.01 & 95     &  & -1.2 & 11   & 11  & 1.04 & 94     \\
        &        & $\beta_3$ & -0.4 & 29   & 28  & 0.99 & 95     &  & 0.2  & 20   & 20  & 1.00 & 95     \\
        & agg111 & $\beta_0$ & 2.9  & 25   & 23  & 1.02 & 94     &  & 1.3  & 17   & 16  & 1.07 & 95     \\
        &        & $\beta_1$ & -0.6 & 28   & 26  & 1.07 & 95     &  & -0.7 & 19   & 18  & 1.10 & 95     \\
        &        & $\beta_2$ & -3.2 & 17   & 16  & 0.96 & 93     &  & -1.2 & 11   & 11  & 1.05 & 94     \\
        &        & $\beta_3$ & -0.5 & 28   & 26  & 1.05 & 94     &  & 0.2  & 19   & 19  & 1.12 & 95     \\
        & agg101 & $\beta_0$ & 2.9  & 25   & 24  & 0.99 & 94     &  & 1.3  & 17   & 17  & 1.05 & 96     \\
        &        & $\beta_1$ & -0.7 & 27   & 26  & 1.08 & 95     &  & -0.7 & 19   & 18  & 1.10 & 94     \\
        &        & $\beta_2$ & -3.2 & 17   & 16  & 0.96 & 94     &  & -1.2 & 11   & 11  & 1.05 & 94     \\
        &        & $\beta_3$ & -0.2 & 29   & 28  & 0.95 & 94     &  & 0.3  & 20   & 20  & 0.99 & 95     \\
        & omn111 & $\beta_0$ & 2.8  & 25   & 24  & 1.02 & 94     &  & 1.3  & 17   & 17  & 1.07 & 95     \\
        &        & $\beta_1$ & -0.8 & 27   & 26  & 1.08 & 95     &  & -0.8 & 19   & 19  & 1.09 & 95     \\
        &        & $\beta_2$ & -3.2 & 16   & 16  & 0.98 & 94     &  & -1.2 & 11   & 11  & 1.05 & 94     \\
        &        & $\beta_3$ & -0.3 & 28   & 27  & 1.02 & 94     &  & 0.3  & 19   & 19  & 1.07 & 96     \\\midrule
$\rho$=0.4 & ave111 & $\beta_0$ & 2.5  & 24   & 24  & 1.05 & 95     &  & 1.0  & 17   & 17  & 1.06 & 95     \\
        &        & $\beta_1$ & -0.3 & 27   & 27  & 1.09 & 95     &  & -0.3 & 19   & 19  & 1.09 & 95     \\
        &        & $\beta_2$ & -2.8 & 16   & 16  & 1.04 & 95     &  & -1.2 & 12   & 11  & 1.05 & 94     \\
        &        & $\beta_3$ & -0.7 & 28   & 27  & 1.06 & 95     &  & 0.2  & 19   & 19  & 1.08 & 95     \\
        & ave101 & $\beta_0$ & 2.6  & 24   & 24  & 1.03 & 95     &  & 1.0  & 17   & 17  & 1.04 & 95     \\
        &        & $\beta_1$ & -0.3 & 27   & 26  & 1.11 & 95     &  & -0.3 & 19   & 18  & 1.11 & 94     \\
        &        & $\beta_2$ & -2.9 & 16   & 16  & 1.04 & 95     &  & -1.2 & 11   & 11  & 1.06 & 94     \\
        &        & $\beta_3$ & -0.6 & 29   & 28  & 0.99 & 95     &  & 0.2  & 20   & 20  & 1.00 & 96     \\
        & agg111 & $\beta_0$ & 2.6  & 24   & 23  & 1.04 & 95     &  & 0.9  & 17   & 16  & 1.07 & 95     \\
        &        & $\beta_1$ & -0.1 & 27   & 26  & 1.10 & 95     &  & -0.2 & 19   & 18  & 1.11 & 94     \\
        &        & $\beta_2$ & -3.0 & 16   & 15  & 0.98 & 94     &  & -1.2 & 12   & 11  & 1.05 & 93     \\
        &        & $\beta_3$ & -0.8 & 27   & 26  & 1.08 & 94     &  & 0.2  & 18   & 18  & 1.16 & 95     \\
        & agg101 & $\beta_0$ & 2.6  & 25   & 24  & 1.00 & 94     &  & 0.9  & 17   & 17  & 1.04 & 95     \\
        &        & $\beta_1$ & -0.3 & 27   & 26  & 1.10 & 94     &  & -0.3 & 19   & 18  & 1.11 & 94     \\
        &        & $\beta_2$ & -3.1 & 16   & 15  & 0.98 & 94     &  & -1.2 & 12   & 11  & 1.05 & 93     \\
        &        & $\beta_3$ & -0.5 & 29   & 28  & 0.95 & 95     &  & 0.2  & 20   & 20  & 0.98 & 96     \\
        & omn111 & $\beta_0$ & 2.6  & 24   & 23  & 1.04 & 95     &  & 0.9  & 17   & 16  & 1.07 & 96     \\
        &        & $\beta_1$ & -0.3 & 27   & 26  & 1.12 & 95     &  & -0.3 & 19   & 18  & 1.12 & 95     \\
        &        & $\beta_2$ & -3.1 & 16   & 15  & 1.00 & 94     &  & -1.2 & 11   & 11  & 1.06 & 94     \\
        &        & $\beta_3$ & -0.6 & 28   & 26  & 1.05 & 95     &  & 0.3  & 19   & 19  & 1.10 & 95     \\\midrule
$\rho$=0.8 & ave111 & $\beta_0$ & 2.4  & 24   & 24  & 1.06 & 95     &  & 0.8  & 17   & 17  & 1.07 & 95     \\
        &        & $\beta_1$ & -0.1 & 27   & 26  & 1.11 & 95     &  & -0.1 & 19   & 18  & 1.10 & 94     \\
        &        & $\beta_2$ & -2.8 & 16   & 16  & 1.05 & 96     &  & -1.0 & 11   & 11  & 1.07 & 94     \\
        &        & $\beta_3$ & -0.6 & 28   & 27  & 1.08 & 95     &  & 0.4  & 19   & 19  & 1.10 & 95     \\
        & ave101 & $\beta_0$ & 2.4  & 25   & 24  & 1.02 & 95     &  & 0.9  & 17   & 17  & 1.04 & 95     \\
        &        & $\beta_1$ & -0.1 & 27   & 26  & 1.12 & 95     &  & -0.1 & 19   & 18  & 1.12 & 94     \\
        &        & $\beta_2$ & -2.8 & 16   & 15  & 1.05 & 95     &  & -1.0 & 11   & 11  & 1.08 & 95     \\
        &        & $\beta_3$ & -0.5 & 29   & 28  & 0.99 & 95     &  & 0.3  & 20   & 20  & 1.00 & 95     \\
        & agg111 & $\beta_0$ & 2.6  & 25   & 23  & 1.04 & 95     &  & 0.6  & 17   & 16  & 1.08 & 95     \\
        &        & $\beta_1$ & 0.2  & 27   & 26  & 1.08 & 93     &  & 0.0  & 19   & 18  & 1.06 & 94     \\
        &        & $\beta_2$ & -2.8 & 17   & 15  & 0.95 & 93     &  & -0.9 & 12   & 11  & 1.02 & 94     \\
        &        & $\beta_3$ & -1.1 & 27   & 26  & 1.12 & 94     &  & 0.3  & 18   & 18  & 1.21 & 95     \\
        & agg101 & $\beta_0$ & 2.5  & 25   & 24  & 0.98 & 95     &  & 0.7  & 17   & 17  & 1.02 & 96     \\
        &        & $\beta_1$ & 0.0  & 27   & 26  & 1.08 & 93     &  & -0.1 & 19   & 18  & 1.07 & 94     \\
        &        & $\beta_2$ & -2.8 & 16   & 15  & 0.95 & 93     &  & -0.9 & 12   & 11  & 1.02 & 93     \\
        &        & $\beta_3$ & -0.5 & 30   & 28  & 0.94 & 94     &  & 0.3  & 20   & 20  & 0.98 & 94     \\
        & omn111 & $\beta_0$ & 2.4  & 25   & 23  & 1.05 & 95     &  & 0.7  & 17   & 16  & 1.08 & 96     \\
        &        & $\beta_1$ & 0.0  & 27   & 26  & 1.13 & 94     &  & -0.2 & 19   & 18  & 1.11 & 94     \\
        &        & $\beta_2$ & -2.9 & 16   & 15  & 1.00 & 94     &  & -1.0 & 12   & 11  & 1.05 & 94     \\
        &        & $\beta_3$ & -0.6 & 28   & 26  & 1.07 & 94     &  & 0.4  & 19   & 18  & 1.12 & 95    \\\bottomrule
\end{tabular}
\end{table}

\begin{table}
\centering
\caption{ \label{tableS5}Overall evaluations for mis-specified working models under the combination of different sample sizes ($n$) and correlations ($\rho$), given three secondary data are partially observed due to informative missingness. MCSE: Monte Carlo standard error; ASE: asymptotic standard error; ERE: empirical relative efficiency; CP: coverage probability. All values (except ERE) are multiplied by $100$.}
\scriptsize
\begin{tabular}{ccccccccccccc}
\toprule
       &                         & \multicolumn{5}{c}{n=300}          &  & \multicolumn{5}{c}{n=600}          \\ \cline{3-7} \cline{9-13} 
       &                         & Bias & MCSE & ASE  & ERE  & 95\%CP &  & Bias & MCSE & ASE  & ERE  & 95\%CP \\ \midrule
ave111 & $\beta_0$ & 2.1  & 24.2 & 23.4 & 1.08 & 96     &  & 0.4  & 16.5 & 16.5 & 1.10 & 95     \\
       & $\beta_1$ & -2.5 & 26.6 & 25.8 & 1.16 & 95     &  & -0.6 & 18.5 & 18.2 & 1.16 & 95     \\
       & $\beta_2$ & -2.1 & 16.3 & 15.3 & 1.10 & 94     &  & -0.6 & 11.1 & 10.8 & 1.09 & 95     \\
       & $\beta_3$ & 1.1  & 27.2 & 26.6 & 1.05 & 95     &  & 1.0  & 19.8 & 18.8 & 1.09 & 95     \\ \midrule
ave101 & $\beta_0$ & 2.3  & 24.4 & 23.5 & 1.06 & 96     &  & 0.6  & 16.6 & 16.6 & 1.08 & 95     \\
       & $\beta_1$ & -2.3 & 26.2 & 25.2 & 1.20 & 95     &  & -0.5 & 18.1 & 17.8 & 1.21 & 95     \\
       & $\beta_2$ & -2.0 & 16.2 & 15.1 & 1.12 & 94     &  & -0.6 & 11.0 & 10.7 & 1.11 & 94     \\
       & $\beta_3$ & 0.6  & 28.3 & 27.6 & 0.98 & 95     &  & 0.6  & 20.8 & 19.5 & 0.99 & 95     \\ \midrule
agg111 & $\beta_0$ & 1.0  & 24.2 & 22.5 & 1.09 & 93     &  & -0.2 & 16.0 & 16.0 & 1.16 & 95     \\
       & $\beta_1$ & -1.5 & 26.1 & 24.6 & 1.21 & 95     &  & -0.4 & 17.8 & 17.4 & 1.26 & 95     \\
       & $\beta_2$ & -2.0 & 16.2 & 14.8 & 1.11 & 93     &  & -0.7 & 11.0 & 10.5 & 1.10 & 94     \\
       & $\beta_3$ & 2.2  & 27.2 & 25.2 & 1.04 & 93     &  & 1.9  & 19.0 & 17.8 & 1.18 & 94     \\ \midrule
agg101 & $\beta_0$ & 2.0  & 24.7 & 23.2 & 1.04 & 94     &  & 0.6  & 16.5 & 16.5 & 1.10 & 95     \\
       & $\beta_1$ & -1.6 & 26.0 & 24.6 & 1.22 & 94     &  & -0.4 & 17.8 & 17.4 & 1.26 & 95     \\
       & $\beta_2$ & -2.1 & 16.2 & 14.8 & 1.11 & 93     &  & -0.7 & 11.0 & 10.5 & 1.10 & 93     \\
       & $\beta_3$ & 0.4  & 29.0 & 27.4 & 0.93 & 94     &  & 0.6  & 21.0 & 19.5 & 0.97 & 94     \\ \midrule
omi111 & $\beta_0$ & 2.0  & 24.2 & 22.9 & 1.09 & 95     &  & 0.7  & 16.2 & 16.2 & 1.13 & 95     \\
       & $\beta_1$ & -2.5 & 25.9 & 24.8 & 1.22 & 94     &  & -1.1 & 17.9 & 17.6 & 1.24 & 95     \\
       & $\beta_2$ & -2.7 & 16.1 & 14.9 & 1.11 & 94     &  & -1.4 & 11.0 & 10.6 & 1.10 & 94     \\
       & $\beta_3$ & 1.3  & 27.4 & 26.0 & 1.03 & 94     &  & 1.3  & 19.7 & 18.4 & 1.10 & 94  \\\bottomrule  
\end{tabular}
\end{table}

\begin{table}
\centering
\caption{\revision{Evaluations for three estimators under the sample sizes $600$ \revisiontwo{where secondary outcomes are very mildly associated with the primary outcome (around $0.15$)}, given three correlation values between two longitudinal secondary outcomes $(\rho=0,0.4,0.8)$. MCSD: Monte Carlo standard deviation; ASE: asymptotic standard error; ERE: empirical relative efficiency (ratio of mean squared errors between the MLE and the target estimate); CP: coverage probability. All
values except for ERE have been multiplied by 100.}}\label{little association}
\footnotesize
\begin{tabular}{lccccccc}
\toprule
       &                         &    Param   & Bias  & MCSD & ASE & ERE  & 95\%CP \\\midrule
ave111 & $\rho=0$   & $\beta_0$ & 0.28  & 17   & 17  & 1.01 & 96     \\
       &                         & $\beta_1$ & -0.25 & 19   & 19  & 1.02 & 95     \\
       &                         & $\beta_2$ & -1.09 & 11   & 11  & 1.03 & 95     \\
       &                         & $\beta_3$ & 0.68  & 19   & 19  & 1.01 & 95     \\
       & $\rho=0.4$ & $\beta_0$ & 0.27  & 17   & 17  & 1    & 95     \\
       &                         & $\beta_1$ & -0.23 & 19   & 19  & 1.03 & 95     \\
       &                         & $\beta_2$ & -1.2  & 12   & 11  & 1.04 & 95     \\
       &                         & $\beta_3$ & 0.60  & 20   & 19  & 1.01 & 95     \\
       & $\rho=0.8$ & $\beta_0$ & 0.32  & 17   & 17  & 1    & 95     \\
       &                         & $\beta_1$ & -0.13 & 19   & 19  & 1.03 & 95     \\
       &                         & $\beta_2$ & -1.09 & 12   & 11  & 1.04 & 95     \\
       &                         & $\beta_3$ & 0.44  & 20   & 19  & 1.01 & 95     \\\midrule
omn111 & $\rho=0$   & $\beta_0$ & 0.21  & 17   & 17  & 1.01 & 96     \\
       &                         & $\beta_1$ & -0.23 & 19   & 19  & 1.02 & 95     \\
       &                         & $\beta_2$ & -1.05 & 11   & 11  & 1.06 & 95     \\
       &                         & $\beta_3$ & 0.83  & 19   & 19  & 1.01 & 95     \\
       & $\rho=0.4$ & $\beta_0$ & 0.19  & 17   & 17  & 1    & 95     \\
       &                         & $\beta_1$ & -0.2  & 19   & 19  & 1.03 & 95     \\
       &                         & $\beta_2$ & -1.16 & 11   & 11  & 1.06 & 95     \\
       &                         & $\beta_3$ & 0.75  & 20   & 19  & 1.01 & 95     \\
       & $\rho=0.8$ & $\beta_0$ & 0.25  & 17   & 17  & 1    & 95     \\
       &                         & $\beta_1$ & -0.11 & 19   & 19  & 1.03 & 95     \\
       &                         & $\beta_2$ & -1.06 & 11   & 11  & 1.07 & 95     \\
       &                         & $\beta_3$ & 0.58  & 20   & 19  & 1.01 & 95     \\\midrule
agg111 & $\rho=0$   & $\beta_0$ & 0.16  & 17   & 17  & 0.99 & 95     \\
       &                         & $\beta_1$ & -0.13 & 19   & 19  & 1.01 & 95     \\
       &                         & $\beta_2$ & -1.04 & 11   & 11  & 1.04 & 95     \\
       &                         & $\beta_3$ & 0.93  & 20   & 19  & 0.99 & 94     \\
       & $\rho=0.4$ & $\beta_0$ & 0.12  & 17   & 17  & 0.98 & 95     \\
       &                         & $\beta_1$ & -0.08 & 20   & 19  & 1.01 & 95     \\
       &                         & $\beta_2$ & -1.14 & 11   & 11  & 1.05 & 95     \\
       &                         & $\beta_3$ & 0.84  & 20   & 19  & 0.99 & 94     \\
       & $\rho=0.8$ & $\beta_0$ & 0.17  & 17   & 17  & 0.96 & 95     \\
       &                         & $\beta_1$ & 0.01  & 20   & 19  & 0.99 & 94     \\
       &                         & $\beta_2$ & -1.05 & 12   & 11  & 1.02 & 95     \\
       &                         & $\beta_3$ & 0.67  & 20   & 19  & 0.98 & 94   \\\bottomrule 
\end{tabular}
\end{table}

\begin{table}
\centering
\caption{Evaluations for MinBo estimators (e.g., ave111 and agg111) under two sample sizes $(n=300, 600)$ with higher correlation between the primary endpoint and secondary outcomes, given correctly specified secondary models (Correct) and mis-specified secondary models (Incorrect). MCSD: Monte Carlo standard deviation; ASE: asymptotic standard error; ERE: empirical relative efficiency (ratio of mean squared errors between the MLE and the target estimate); CP: coverage probability based on 95\% confidence internal. All
values except for ERE have been multiplied by 100.}\label{high cor}
\begin{tabular}{llccccccccccc}
\toprule
       &                         & \multicolumn{5}{c}{n=300}          &  & \multicolumn{5}{c}{n=600}          \\ \cline{3-7} \cline{9-13} 
       &                         & Bias  & MCSD & ASE & ERE  & CP &  & Bias  & MCSD & ASE & ERE  & CP \\\midrule
ave111 & $\beta_0$ & 2.20  & 22   & 22  & 1.21 & 95     &  & 1.13  & 15   & 15  & 1.20 & 95     \\
Correct     & $\beta_1$ & -1.31 & 23   & 22  & 1.46 & 94     &  & -0.68 & 16   & 16  & 1.47 & 94     \\
       & $\beta_2$ & -2.02 & 14   & 13  & 1.52 & 93     &  & -0.93 & 10   & 9   & 1.59 & 94     \\
       & $\beta_3$ & 1.31  & 24   & 24  & 1.31 & 95     &  & 0.37  & 17   & 17  & 1.32 & 94     \\\midrule
agg111 & $\beta_0$ & 1.46  & 23   & 21  & 1.17 & 94     &  & 0.52  & 15   & 15  & 1.19 & 95     \\
Correct      & $\beta_1$ & -0.84 & 25   & 23  & 1.33 & 94     &  & -0.10 & 17   & 16  & 1.38 & 96     \\
       & $\beta_2$ & -0.84 & 15   & 15  & 1.34 & 93     &  & -0.26 & 10   & 10  & 1.33 & 95     \\
       & $\beta_3$ & 1.25  & 22   & 21  & 1.47 & 94     &  & 0.29  & 15   & 15  & 1.67 & 94     \\ \midrule
ave111 & $\beta_0$ & 2.52  & 23   & 23  & 1.07 & 95     &  & 1.18  & 16   & 16  & 1.07 & 95     \\
Incorrect     & $\beta_1$ & -1.75 & 26   & 24  & 1.23 & 94     &  & -0.71 & 17   & 16  & 1.36 & 94     \\
       & $\beta_2$ & -2.64 & 16   & 14  & 1.19 & 92     &  & -1.16 & 11   & 10  & 1.28 & 94     \\
       & $\beta_3$ & 1.52  & 27   & 28  & 0.99 & 95     &  & 0.44  & 20   & 20  & 1.00 & 95     \\ \midrule
agg111 & $\beta_0$ & 2.48  & 22   & 21  & 1.20 & 93     &  & 0.97  & 15   & 14  & 1.29 & 95     \\
Incorrect    & $\beta_1$ & -1.86 & 25   & 21  & 1.32 & 93     &  & -0.46 & 15   & 15  & 1.61 & 94     \\
       & $\beta_2$ & -1.98 & 14   & 13  & 1.51 & 92     &  & -0.79 & 9    & 9   & 1.73 & 94     \\
       & $\beta_3$ & 1.58  & 22   & 21  & 1.58 & 95     &  & 0.54  & 15   & 15  & 1.71 & 94  \\\bottomrule  
\end{tabular}
\end{table}

\begin{table}
\centering
\caption{\revision{Evaluations for three estimators (\revisiontwo{ave11, agg11, joint likelihood (JL)}) under two sample sizes $(n=300, 600)$, given correctly specified secondary models (Correct) and mis-specified secondary models (Incorrect). MCSD: Monte Carlo standard deviation; ASE: asymptotic standard error; ERE: empirical relative efficiency (ratio of mean squared errors between the MLE and the target estimate). All
values except for ERE have been multiplied by 100.}}\label{jl}
\footnotesize
\begin{tabular}{cccccccccc}
\toprule
       &     &           & \multicolumn{3}{c}{Correct} &  & \multicolumn{3}{c}{Incorrect} \\ \cline{4-6} \cline{8-10} 
Method & Sample Size   & Param & Bias     & MCSD    & ERE    &  & Bias     & MCSD     & ERE     \\\midrule
ave11  & 300 & $\beta_0$   & 0.36     & 8.69    & 1.41   &  & 0.45     & 9.50     & 1.18    \\
       &     & $\beta_1$   & -0.32    & 8.88    & 1.94   &  & -0.48    & 10.5     & 1.38    \\
       &     & $\beta_2$   & 0.19     & 4.28    & 2.06   &  & 0.17     & 4.71     & 1.71    \\
       &     & $\beta_3$   & -0.32    & 8.86    & 1.62   &  & -0.3     & 10.58    & 1.14    \\
       & 600 & $\beta_0$   & 0.16     & 5.72    & 1.47   &  & 0.22     & 6.34     & 1.2     \\
       &     & $\beta_1$   & -0.09    & 6.06    & 1.99   &  & -0.2     & 7.21     & 1.4     \\
       &     & $\beta_2$   & 0.08     & 2.82    & 2.25   &  & 0.08     & 3.13     & 1.83    \\
       &     & $\beta_3$   & -0.35    & 6.16    & 1.74   &  & -0.37    & 7.50     & 1.17    \\\midrule
agg11  & 300 & $\beta_0$   & 0.26     & 9.94    & 1.08   &  & 0.49     & 9.30     & 1.24    \\
       &     & $\beta_1$   & -0.25    & 11.29   & 1.2    &  & -0.75    & 10.51    & 1.38    \\
       &     & $\beta_2$   & 0.11     & 5.19    & 1.4    &  & 0.15     & 4.85     & 1.61    \\
       &     & $\beta_3$   & -0.18    & 10.94   & 1.06   &  & -0.11    & 10.34    & 1.19    \\
       & 600 & $\beta_0$   & 0.03     & 6.55    & 1.12   &  & 0.2      & 6.10     & 1.29    \\
       &     & $\beta_1$   & 0.10     & 7.65    & 1.25   &  & -0.27    & 7.11     & 1.44    \\
       &     & $\beta_2$   & 0.01     & 3.44    & 1.52   &  & 0.01     & 3.27     & 1.67    \\
       &     & $\beta_3$   & -0.26    & 7.18    & 1.28   &  & -0.28    & 7.05     & 1.33    \\\midrule
JL     & 300 & $\beta_0$   & 0.24     & 6.59    & 2.46   &  & 30.36    & 7.99     & 0.11    \\
       &     & $\beta_1$   & -0.22    & 4.3     & 8.24   &  & 37.05    & 9.61     & 0.01    \\
       &     & $\beta_2$   & 0.04     & 2.24    & 7.58   &  & 0.01     & 3.72     & 2.74    \\
       &     & $\beta_3$   & -0.10    & 3.87    & 8.49   &  & 76.22    & 8.07     & 0.02    \\
       & 600 & $\beta_0$   & 0.07     & 4.52    & 2.36   &  & 30.44    & 5.48     & 0.05    \\
       &     & $\beta_1$   & -0.17    & 3.14    & 7.37   &  & 37.23    & 6.22     & 0       \\
       &     & $\beta_2$   & 0.05     & 1.53    & 7.65   &  & 0.03     & 2.73     & 2.4     \\
       &     & $\beta_3$   & -0.08    & 2.74    & 8.8    &  & 76.26    & 5.08     & 0.01   \\\bottomrule
\end{tabular}
\end{table}